\documentclass[lettersize,journal]{IEEEtran}
\usepackage{amssymb}
\usepackage{amsmath,amsfonts}
\usepackage{amsthm}
\usepackage{algorithmic}
\usepackage{algorithm}
\usepackage{multirow}
\usepackage{mathrsfs}
\usepackage{array}
\usepackage{textcomp}
\usepackage{stfloats}
\usepackage{float}
\usepackage{booktabs}
\usepackage{verbatim}
\usepackage{url}
\usepackage{graphicx}
\usepackage{subfigure}
\usepackage{cite}
\begin{document}

\title{Federated Learning Strategies for Coordinated Beamforming in Multicell ISAC}
\author{Lai Jiang,~\IEEEmembership{Student Member,~IEEE,} Kaitao Meng,~\IEEEmembership{Member,~IEEE,} Murat Temiz,~\IEEEmembership{Member,~IEEE,} Jiaming Hu,~\IEEEmembership{Student Member,~IEEE,} and Christos Masouros,~\IEEEmembership{Fellow,~IEEE,}
        
\thanks{The authors are with the Department of Electronic and Electrical Engineering, University College London, London, UK (emails: \{lai.jiang.22, kaitao.meng, m.temiz, jiaming.hu.19, c.masouros\}\@ucl.ac.uk).}}

\IEEEpubid{0000--0000/00\$00.00~\copyright~2021 IEEE}

\maketitle

\begin{abstract}
We propose two cooperative beamforming frameworks based on federated learning (FL) for multi-cell integrated sensing and communications (ISAC) systems. Our objective is to address the following dilemma in multicell ISAC: 1) Beamforming strategies that rely solely on local channel information risk generating significant inter-cell interference (ICI), which degrades network performance for both communication users and sensing receivers in neighboring cells; 2) conversely centralized beamforming strategies can mitigate ICI by leveraging global channel information, but they come with substantial transmission overhead and latency that can be prohibitive for latency-sensitive and source-constrained applications. To tackle these challenges, we first propose a partially decentralized training framework motivated by the vertical federated learning (VFL) paradigm. In this framework, the participating base stations (BSs) collaboratively design beamforming matrices under the guidance of a central server. The central server aggregates local information from the BSs and provides feedback, allowing BSs to implicitly manage ICI without accessing the global channel information. To make the solution scalable for densely deployed wireless networks, we take further steps to reduce communication overhead by presenting a fully decentralized design based on the horizontal federated learning (HFL). Specifically, we develop a novel loss function to control the interference leakage power, enabling a more efficient training process by entirely eliminating local channel information exchange. Numerical results show that the proposed solutions can achieve significant performance improvements comparable to the benchmarks in terms of both communication and radar information rates.
\end{abstract}


\section{Introduction}
Integrated sensing and communications (ISAC) has emerged as a key technology for the next-generation wireless networks \cite{liu2020joint, zhang2021overview}. As a promising solution to address the challenge of limited wireless resources, ISAC systems unify both radar and communication functions on a shared spectrum, waveform, and platform, enabling more efficient and innovative use of available resources. In addition, by simultaneously integrating sensing and reliable communication service, ISAC paves the way for advanced applications such as autonomous vehicles, smart cities, and enhanced security systems, making it a cornerstone technology for future networks \cite{liu2022integrated}. In the cellular domain, the ultimate goal of deploying ISAC solutions would be to enable coordinated sensing of unprecedented scale. This will inevitably require network level coordination to address interference and enable efficient use of resources \cite{meng2024cooperative2}.

With the deployment of multi-antenna arrays on the base stations (BSs), transmit beamforming becomes a critical technique in the downlink transmission, as it can enable flexible tradeoffs and mutual benefits between the communication and sensing by leveraging the available spatial degrees of freedom (DoFs) \cite{meng2023sensing}. Numerous works have studied transmit waveforms/beamforming design to enhance both S\&C performance at the link level \cite{liu2018toward, liuxiang2020joint, dong2020low, liu2021cramer, Hua10086626, li2024framework}, focusing on optimizing the communication and sensing performance for users and targets within a single cell. However, in practical multi-cell systems where multiple BSs may share the same time and frequency resources, users experience not only intra-cell interference caused from unintended signals in the same cell but also inter-cell interference (ICI) from surrounding cells, which results in reduced received signal-to-interference-plus-noise ratio (SINR). In fact, ICI has been proven detrimental to both network-level S\&C performance in dense deployed cellular networks \cite{meng2023network}, with sensing being particularly sensitive due to the round-trip path loss of echo signals. Consequently, per-cell beamforming strategies that overlook the effect of ICI can not be directly applied in practical multi-cell ISAC systems.

\IEEEpubidadjcol
A potential solution for ICI management in cellular systems is to either enable distributed coordinated beamforming by exchanging local information between multiple BSs or to gather global information for designing the optimal transmit waveform or beamforming matrices in a centralized manner \cite{meng2024cooperative1}. Specifically, the authors in \cite{meng2024cooperative2} proposed a framework where BSs cooperatively serve the users and localize each target for enhancing the ICI management and S\&C performance. Moreover, in \cite{babu2024precoding}, a coordinated beamforming (CBF) framework is developed to optimize target parameter estimation and communication performance while managing the ICI power, where precoders are centrally designed and subsequently allocated to individual BSs for local execution. Nevertheless, the strategies that rely on information exchange inevitably incur excessive transmission overhead and latency, particularly in the large-scale densely deployed networks. Therefore, it is crucial to develop decentralized beamforming strategies that can effectively manage ICI while minimizing the need for frequent information exchange.

In addition to minimizing communication overhead, computational complexity is a significant concern in the realm of beamforming design \cite{qian2014low}. In the literature, the majority of existing beamforming methods, whether for communication-only systems or ISAC systems\cite{liu2018toward, liuxiang2020joint, dong2020low, liu2021cramer, Hua10086626, li2024framework, shi2011iteratively} are developed based on optimization techniques. For instance, the authors in \cite{liu2018toward, Hua10086626} formulated the optimization problem with the objective of maximizing the radar sensing performance while guaranteeing the SINR for each communication user exceeds a predefined threshold, and the optimal beamforming matrices can then be obtained by solving the problem using semidefinite relaxation (SDR) technique. In \cite{peng2024mutual}, the authors extended the WMMSE \cite{shi2011iteratively} framework to ISAC systems to address the proposed problem of maximizing the weighted S\&C sum rate. However, optimization-based algorithms often rely heavily on iterative procedures and complex matrix operations, which introduce substantial computational complexity and latency \cite{elbir2023twenty}. This makes the implementation of these algorithms in real-world systems challenging, particularly in time-sensitive scenarios or those with limited processing resources.

\IEEEpubidadjcol
Building on the above discussion, deep learning techniques provide an efficient alternative to tackle the optimization problems through a data-driven strategy \cite{sun2018learning, xia2019deep, huang2018unsupervised, lin2019beamforming, hojatian2021unsupervised}. The deep neural networks (DNNs) exhibit a remarkable capacity for automatically extracting features from large amounts of raw data and gradually abstracting high-level representations through multiple layers of interconnected nodes or computational units \cite{jia2016deep}. In contrast to methods based on rigorous mathematical models, an offline-trained DNN can efficiently handle large-scale matrix multiplications and additions, with the exploitation of its parallel computation capabilities. 
Up to now, extensive studies have investigated the application of DNN to find the optimal beamformer/precoder using channel state information (CSI) for both conventional communication-only systems and ISAC systems. For example, the authors in \cite{xia2019deep} proposed a DNN-based framework for downlink beamforming, training the network to minimize the mean square error (MSE) between the output and the desired beamforming matrix. Furthermore, to avoid excessive computation required in supervised learning methods, the authors in \cite{huang2018unsupervised} introduced an unsupervised approach, training the DNN to directly maximize the sum rate under transmit power constraints, which can achieve the performance comparable to WMMSE algorithm \cite{shi2011iteratively}. In \cite{sankar2023learning}, an neural network was developed for ISAC precoders design, aiming to maximize target illumination power while maintaining the quality of service (QoS) for communication users. Despite achieving low computational complexity, the aforementioned references often fall short in effectively managing ICI. This is primarily due to their reliance on per-cell beamforming strategies, which overlook the effects of ICI from adjacent cells. 

The above discussion motivates us to adopt the federated learning (FL) paradigm for beamforming design in multi-cell ISAC systems, as it provides a framework that integrates distributed local information while minimizing the overhead associated with information exchange. As a paradigm of distributed learning framework, FL enables multiple local clients to collaboratively train a DNN model under coordination of a central server \cite{mcmahan2017communication}. 
Unlike centralized learning schemes, during the training process in FL, each client only transmits the gradients or model parameters, rather than the entire dataset. This significantly reduces the transmission overhead, while at the same time preserving privacy. Thus, the jointly trained model leverages the datasets spread across numerous local nodes without explicitly exchanging them, which aligns well with the needs for transmission/computation efficiency in practical multi-cell systems \cite{chen2021distributed}. To leverage the decentralized nature of FL, recent research efforts are applying FL framework for the physical layer design in massive MIMO systems \cite{elbir2020federated, elbir2021federated, elbir2021federatedforphysical, yi2024model}. Specifically, the authors in \cite{elbir2020federated} applied the FL framework for hybrid beamforming design using the gradients data collected from distributed edge users, which achieved the performance comparable to that of the centralized learning scheme \cite{huang2019fast}. Subsequently in \cite{wang2023federated}, the authors first investigated a decentralized precoding framework for cell-free systems based on horizontal federated learning (HFL). The results showed that integrating different channel datasets with the same feature space can enhance the model performance. Subsequently, they proposed a vertical federated learning (VFL)-based approach to integrate the feature-level information at local devices, which eliminates the communication cost associated with exchanging local CSI during the online precoding stage. Notably, to the best of our knowledge, there are limited research focused on developing FL-based beamforming schemes for multi-cell ISAC systems.

\IEEEpubidadjcol
In this work, our primary objective is to develop an efficient decentralized beamforming method tailored for multi-cell ISAC systems. By leveraging the distributed nature of FL, we propose two coordinated beamforming frameworks that utilize channel data collected from distributed BSs to jointly train a DNN. 
The first VFL-based solution successfully eliminates the need to exchange local channel information during online deployment, enabling the participating BSs to achieve cooperative beamforming design. To enhance scalability and adaptability for practical deployment, we propose a second solution that integrates the concept of HFL with a novel loss function for controlling leakage power. This approach achieves global interference control using only local channel information, enabling fully distributed beamforming during both training and online deployment. The main contributions of this paper are summarized as follows:
\begin{itemize}
	\item We propose a partially decentralized beamforming framework based on VFL. During the offline training phase, multiple dual-functional BSs use their local datasets to train the individual DNN models, guided by the global loss function that is computed by a central server. The trained models then design the optimal beamforming matrices with low computational overhead when deployed online.
    \item We propose a fully decentralized beamforming framework based on HFL. In this framework, we propose a novel loss function to enable a fully decentralized implementation, which aims at maximizing the weighted sum of communication rate and radar information rate while mitigating interference leakage to undesired receivers. Therefore, the loss function can be independently optimized at each BS, which eliminates the need for exchanging channel information during both training and online beamforming stages. This characteristic enhances data privacy and significantly reduces communication overhead, making the approach more adaptable and efficient for multi-cell systems.
	\item Through numerical simulations, we validate the effectiveness of our proposed approaches in eliminating ICI by comparing them with traditional per-cell beamforming strategies. The results highlight the performance improvements achieved by effectively controlling ICI in multi-cell systems. Moreover, our solutions can achieve performance comparable to those of centralized methods, and strike a good tradeoff between communications and sensing performance. Finally, we apply the model pruning technique to further reduce the computational complexity of proposed methods, offering substantial improvements in efficiency and feasibility for practical deployments.
\end{itemize}
\textbf{Notations:} We use the following mathematical notations throughout this paper. Boldface upper-case letters denote the matrices, boldface lower-case letters denote the vectors, and normal font denotes the scalars. $(\mathbf{\cdot})^T$, $(\mathbf{\cdot})^H$, $\Re(\mathbf{\cdot}), \Im(\mathbf{\cdot})$ are transpose, conjugate transpose, the real part and the imaginary part of a matrix/vector respectively. tr$(\mathbf{\cdot})$ is the trace of a matrix and $\Vert \mathbf{\cdot} \Vert$ refers to the Euclidean norm. The complex space and real space are represented by $\mathbb{C}$ and $\mathbb{R}$. $I_K$ is the identity matrix of rank $K$.

\begin{figure}[!t]
\centering
\includegraphics[scale=0.10]{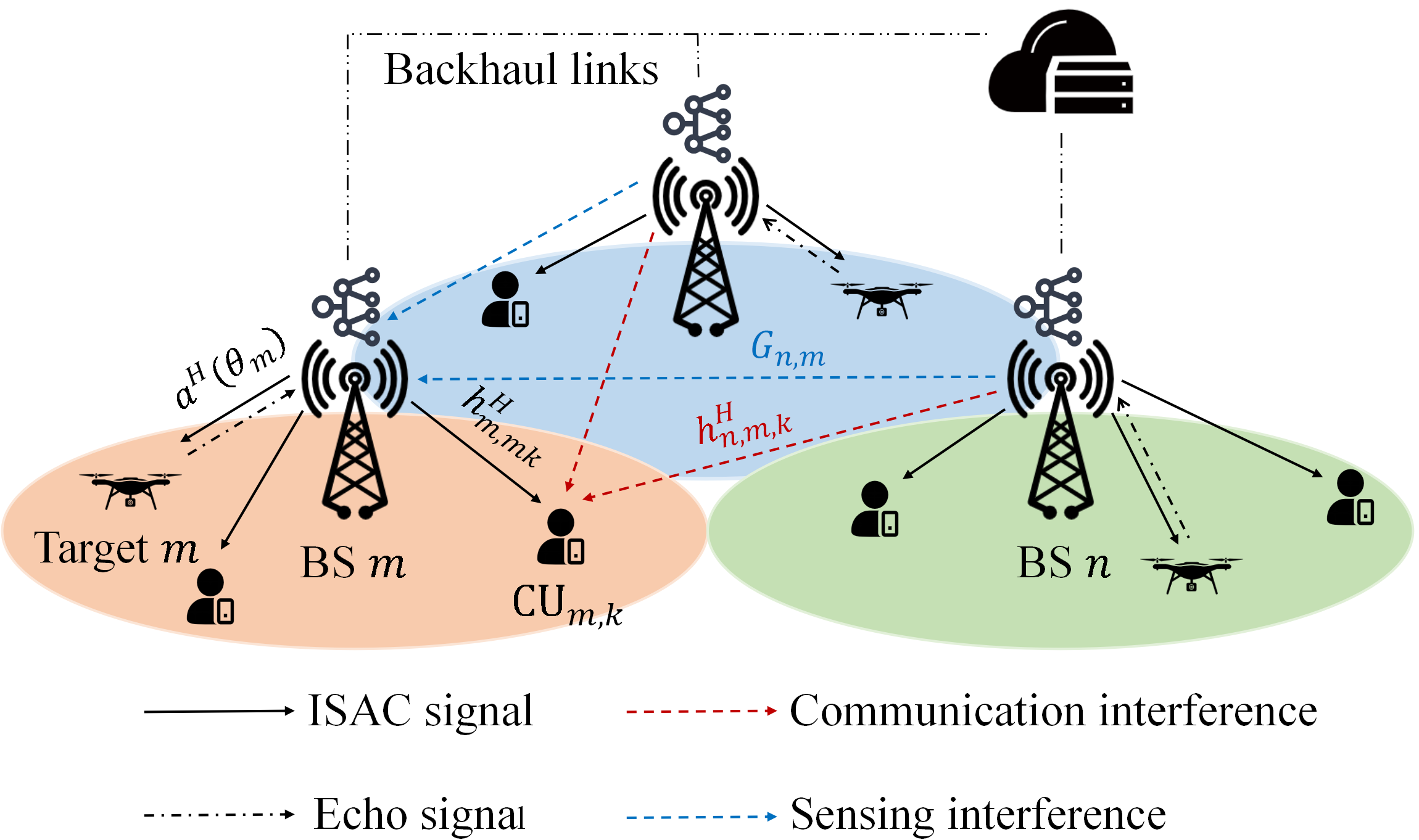}
\caption{The considered multi-cell ISAC system.}
\label{fig_system}
\end{figure}

\section{system model}
\IEEEpubidadjcol
As shown in Fig. \ref{fig_system}, the proposed framework is carried out by $M$ BSs and a central server. The dual-functional BS $m$, $m=1,2,..,M$ is equipped with a half-wavelength spaced uniform linear array (ULA) of $N_T$ transmit antennas and $N_R$ receive antennas, serving $K_m$ downlink single-antenna users while sensing a target simultaneously. During the training stage, the model parameters are exchanged via the backhual links between the participating BSs and the server. The index sets of the communication users and BSs are denoted by $\mathcal{K}_m \stackrel{\triangle}{=} [1,2,...,K_m]$ and $\mathcal{M} \stackrel{\triangle}{=} [1,2,...,M]$, respectively.

\subsection{Communication model}
In the downlink, a unit-power data stream $\mathbf{S}_m \in \mathbb{C}^{K_m \times L}$ with $L$ being the number of timeslots in a frame is transmitted by the $m$th BS to the $K_m$ communication users (CUs) which are uniformly distributed within the coverage area of each cell. Without loss of generality, we assume $K_m=K$, $m=1,2,...,M$ for consistency. The baseband transmitted symbol matrix $\mathbf{X}_m \in \mathbb{C}^{N_T \times L}$ at the $m$th BS is denoted by
\vspace{-1.5mm}
\begin{equation}
    \begin{aligned}\label{eq:x}
\mathbf{X}_m=\mathbf{W}_m\mathbf{S}_m=\sum_{k=1}^{K}\mathbf{w}_{m,k}\mathbf{s}_{m,k},
    \end{aligned}
    \vspace{-1.5mm}
\end{equation}
where $\mathbf{W}_m=[\mathbf{w}_{m,1}, \mathbf{w}_{m,2}, ...\mathbf{w}_{m,K}] \in \mathbb{C}^{N_T \times K}$ is the beamforming matrix to be designed, and $\mathbf{s}^H_{m,k} \in \mathbb{C}^{L \times 1}$ denotes the data stream intended for the $k$th user in the $m$th cell. We assume the data streams are orthogonal when $L$ is sufficiently large so that: $(1/L)\mathbf{S}_m\mathbf{S}^H_m=I_K$. In this case, the received signal at the $k$th user in the $m$th cell is the summation of the intended signal and both intra-cell and inter-cell interference, given by
\vspace{-2.5mm}
\begin{equation}
    \begin{aligned}\label{eq:yc}
        \mathbf{y}^c_{m,k} &= \underbrace{\mathbf{h}^H_{m,m,k}\mathbf{w}_{m,k}\mathbf{s}_{m,k}}_{\rm intended\ signal}+\underbrace{\sum_{l\neq k}^{K}\mathbf{h}^H_{m,m,k}\mathbf{w}_{m,l}\mathbf{s}_{m,l}}_{\rm intra-cell\ interference} \\
        &+\underbrace{\sum_{n\neq m}^{M}\sum_{i=1}^{K}\mathbf{h}^H_{n,m,k}\mathbf{w}_{n,i}\mathbf{s}_{n,i}}_{\rm inter-cell\ interference} + \mathbf{z}_{m,k},
    \end{aligned}
    \vspace{-1.5mm}
\end{equation}
where $\mathbf{h}_{m,n,k} \in \mathbb{C}^{N_T \times 1}$ denotes the block fading channel vector from the $m$th BS to the $k$th user in the $n$th cell, and $\mathbf{z}_{m,k}$ is the additive white Gaussian noise (AWGN) vector, i.e., each element is independent and identically distributed and follows the Gaussian distribution with zero mean and variance $\sigma^2_c$. The received SINR of user $k$ is

\vspace{-1.5mm}
\begin{small}
\begin{equation}
    \begin{aligned}\label{eq:snrc}
    \gamma^c_{m,k} &= \frac{|\mathbf{h}^H_{m,m,k}\mathbf{w}_{m,k}|^2}{\sum_{l\neq k}^{K}|\mathbf{h}^H_{m,m,k}\mathbf{w}_{m,l}|^2 + \sum_{n\neq m}^{M}\sum_{i=1}^{K}|\mathbf{h}^H_{n,m,k}\mathbf{w}_{n,i}|^2 + \sigma^2_c}.
    \end{aligned}
\end{equation}
\end{small}

\IEEEpubidadjcol
The first two terms in the denominator of (\ref{eq:snrc}) represent the power of intra-cell interference and inter-cell interference, respectively, which are to be minimized to improve the system-level performance. Then, the achievable sum communication rate of the multicell ISAC is written as
\vspace{-1.5mm}
\begin{equation}
    \begin{aligned}\label{eq:ssr}
    R_c &= \sum_{m=1}^{M}\sum_{k=1}^{K}\log_{2}(1+\gamma^c_{m,k}).
    \end{aligned}
    \vspace{-1.5mm}
\end{equation}

\subsection{Sensing model}
As commonly adopted in the literature \cite{liu2021cramer, demirhan2023cell, babu2024precoding}, we consider a point target located in the far field of each cell and sensed by its nearest BS. For multi-target scenarios, the BS serves a single target at a given moment to maximize the utilization of its transmission power and avoid sensing intra-cell interference between the targets. As the BS works as a mono-static radar mode, the angle of departure (AoD) and angle of arrival (AoA) are the same. Following the network-level sensing interference model in \cite{meng2023network}, it is the ICI channels from neighboring BSs to the serving BS that impact the reception of the target echoes and therefore dominate the network's sensing performance. The received signal at the $m$th serving BS is the summation of the echo signal reflected by the target and ICI from the surrounding BSs, given by
\begin{equation}
    \begin{aligned}\label{eq:ys}
        \mathbf{y}^s_m &= \underbrace{\alpha_m\mathbf{b}(\theta_m)\mathbf{a}^H(\theta_m)\mathbf{W}_m\mathbf{s}_m(t-2\tau_m)}_{\rm target\ echo\ signal}\\
        &+ \underbrace{\sum_{n\neq m}^{M}\mathbf{G}_{n,m}\mathbf{W}_n\mathbf{s}_n(t-\tau_{n,m})}_{\rm inter-cell\ inteference}\\
        &+ \underbrace{\mathbf{G}_{m,m}\mathbf{W}_m\mathbf{s}_m(t)}_{\rm self-interference} + \mathbf{z}_m,
    \end{aligned} 
\end{equation}
where $\mathbf{G}_{n,m} \in \mathbb{C}^{N_R \times N_T}$ represents the interference channel from the $n$th BS to the $m$th BS, and $\mathbf{G}_{m,m}$ is the self-interference channel from the transmit antennas to the receive antennas of BS $m$. In (\ref{eq:ys}), $\alpha_m$ models the round-trip pathloss and the radar cross section (RCS) of the target, $\tau$ is the delay experienced by the signal. The transmit and receive steering vectors are represented by $\mathbf{a}(\theta_m)=[1,...,e^{j\pi(N_T-1)\sin(\theta_m)}]^T \in\mathbb{C}^{N_T\times 1}$ and $\mathbf{b}(\theta_m)=[1,...,e^{j\pi(N_R-1)\sin(\theta_m)}]^T \in \mathbb{C}^{N_R\times1}$ respectively, and $\theta_m$ denotes the angle of the target in the $m$th cell with respect to its serving BS, which is assumed to be estimated from a previous crude observation. To maximize the received SINR and ensure computational efficiency, the $m$th BS applies the maximum-ratio combining (MRC) beamformer $\mathbf{v}^H_m=\mathbf{b}^H(\theta_m) \in \mathbb{C}^{1\times N_R}$ to process the received signal, the echo signal after processing is given by
\begin{equation}
    \begin{aligned}\label{eq:yseq}
        \mathbf{\tilde{y}}^s_m &= \mathbf{v}^H_m\mathbf{y}^s_m \\
        &= N_R\alpha_m\mathbf{a}^H(\theta_m)\mathbf{W}_m\mathbf{s}_m(t-2\tau_m) \\
        &+ \sum_{n\neq m}^{M}\mathbf{g}^H_{n,m}\mathbf{W}_n\mathbf{s}_n(t-\tau_{n,m}) \\
        &+ \mathbf{v}^H_m\mathbf{G}_{m,m}\mathbf{W}_m\mathbf{s}_m(t) +\mathbf{\tilde{z}}_m,
    \end{aligned}
\end{equation}
where $\mathbf{g}^H_{n,m}=\mathbf{v}^H_m\mathbf{G}_{n,m} \in \mathbb{C}^{1\times N_T}$ is the equivalent interference channel, and $\tilde{z}_m$ is the equivalent AWGN with variance $\sigma^2_s$. As $\mathbf{G}_{m,m}$ can be obtained from previous measurements, we assume the sensing self-interference can be canceled by BS $m$ as in \cite{meng2023network}. Let $\mathbf{g}^H_{m,m}=\alpha_m\mathbf{a}^H(\theta_m)$ denote the equivalent sensing channel from the $m$th BS to the intended target, the SINR of received signal at the $m$th BS can be denoted by
\begin{equation}
    \begin{aligned}\label{eq:snrs}
    \gamma^s_m &=  N_R\frac{\sum_{k=1}^{K}|\mathbf{g}^H_{m,m}\mathbf{w}_{m,k}|^2}{\sum_{n\neq m}^{M}\sum_{l=1}^{K}|\mathbf{g}^H_{n,m}\mathbf{w}_{n,l}|^2 + \sigma^2_s},
    \end{aligned}
\end{equation}
\IEEEpubidadjcol
In (\ref{eq:snrs}), the numerator represents the illumination power for the intended target, which is expected to be maximized for achieving better sensing performance \cite{stoica2007probing}. The achievable sum radar information rate is used to evaluate the network-level sensing performance. Indeed, the accuracy of parameter estimation is proportional to the information rate \cite{yang2007mimo}, which is given by
\begin{equation}
    \begin{aligned}\label{eq:rir}
    R_s &= \sum_{m=1}^{M}\log_{2}(1+\gamma^s_{m}).
    \end{aligned}
\end{equation}

\subsection{Problem Description}
To achieve the performance tradeoff between communications and sensing, we aim to solve the following global optimization problem
\begin{equation}
    \begin{aligned}\label{P:P1}
        &\max_{\mathbf{[W_1,..,W_M]}} \quad \rho R_c + (1-\rho)R_s \\
        &\begin{array}{r@{\quad}r@{}l@{\quad}l}
        s.t. &\text{tr}(\mathbf{W}_m\mathbf{W}^H_m) \leq P_T,\quad \forall m,\\
        \end{array}
    \end{aligned}
\end{equation}
where $\rho \in [0,1]$ is the weighting factor to select between the communication metric and sensing metric, and $P_T$ denotes the transmit power constraint at each BS. Apparently, to maximize the S\&C performance from the network perspective, we need to eliminate the ICI to unintended receivers. However, finding the optimal solution of problem (\ref{P:P1}) is challenging not only due to its non-convexity but also because it requires the knowledge of global channel information, e.g. the communication channels and sensing channels between all the transceivers across the interfering cells. To address the problem mentioned above, we propose two beamforming frameworks with different levels of decentralization and coordination to solve the problem (\ref{P:P1}) based on the VFL and HFL frameworks respectively in Section \ref{VFL} and \ref{HFL}.

\IEEEpubidadjcol
\section{VFL-based Beamforming framework}
\label{VFL}
In the scenario considered, the knowledge of each BS is limited to the channels between itself and all the receivers, i.e., the $m$th BS has the information of local communication channels $\mathbf{H}_m=\{\mathbf{h}_{m,n,k}\}_{n \in \mathcal{M},k \in \mathcal{K}}$ and local sensing channels $\mathbf{G}_m=\{\mathbf{g}_{m,n}\}_{n \in \mathcal{M}}$. 
The global communication and sensing channel information is defined as $\mathbf{H}=[\mathbf{H}_1, \mathbf{H}_2,...\mathbf{H}_M] \in \mathbb{C}^{N_T \times M^2K}$ and $\mathbf{G}=[\mathbf{G}_1, \mathbf{G}_2,...\mathbf{G}_M] \in \mathbb{C}^{N_T \times M^2}$, which can be viewed as the aggregation of the local channel information. In other words, the channel datasets collected by the BSs share the same sample space, consisting of the the $MK$ users and $M$ targets in the network, but each BS possesses distinct local channel information related to these users/targets. This aligns well with the application scenario of the VFL framework, which is designed for situations where partial channel information is available but full global channel information is inaccessible. Leveraging its nature of feature aggregation, we propose using the VFL framework to train local DNNs for the coordinated design of optimal dual-functional beamforming matrices. 


\subsection{Training Samples Generation}
\label{data_collect}
Before the training phase, we introduce a channel sample collection stage where each BS collects its own local channel information to prepare the training datasets. This step is to ensure that the BSs have sufficient data for training the decentralized beamforming framework. Without loss of generality, we assume a time-division duplex (TDD) system so that CSI can be estimated via pilot symbols. First, each BS transmits an orthogonal pilot signal $\tilde{\mathbf{X}}$ for channel estimation and initial detection, the covariance matrix of the probing signal is given by:
\begin{equation}
    \label{eq:cov}
    \mathbf{R}_{\tilde{\mathbf{X}}}=\frac{P_T}{N_T}\mathbf{I}_{N_T}.
\end{equation}

The omnidirectional signals are received by both intended users and targets, as well as by interference devices. BS $m$ can obtain a first crude estimate of the direction $\theta_m$ of the target in its cell based on the echo signal. Once received the pilots, the users and neighboring BSs feed the estimated channel data back to the transmitting BS. Each BS gathers the received channel information and compiles it to form a single channel sample. Finally, BS $m$ obtains a collection of channel samples $\mathbf{H}_m$ and $\mathbf{G}_m$, and the overall local dataset is represented by $\mathcal{D}^{(m)}=\big \{\mathbf{H}_m, \mathbf{G}_m\big \}$. Specifically, it is reasonable to assume that BSs can obtain accurate channel estimation as DNNs have strong robustness toward input data as in \cite{lin2019beamforming, wang2023federated}.

\subsection{DNN architecture}
\begin{figure}[htbp]
\centering
\includegraphics[width=1.0\linewidth]{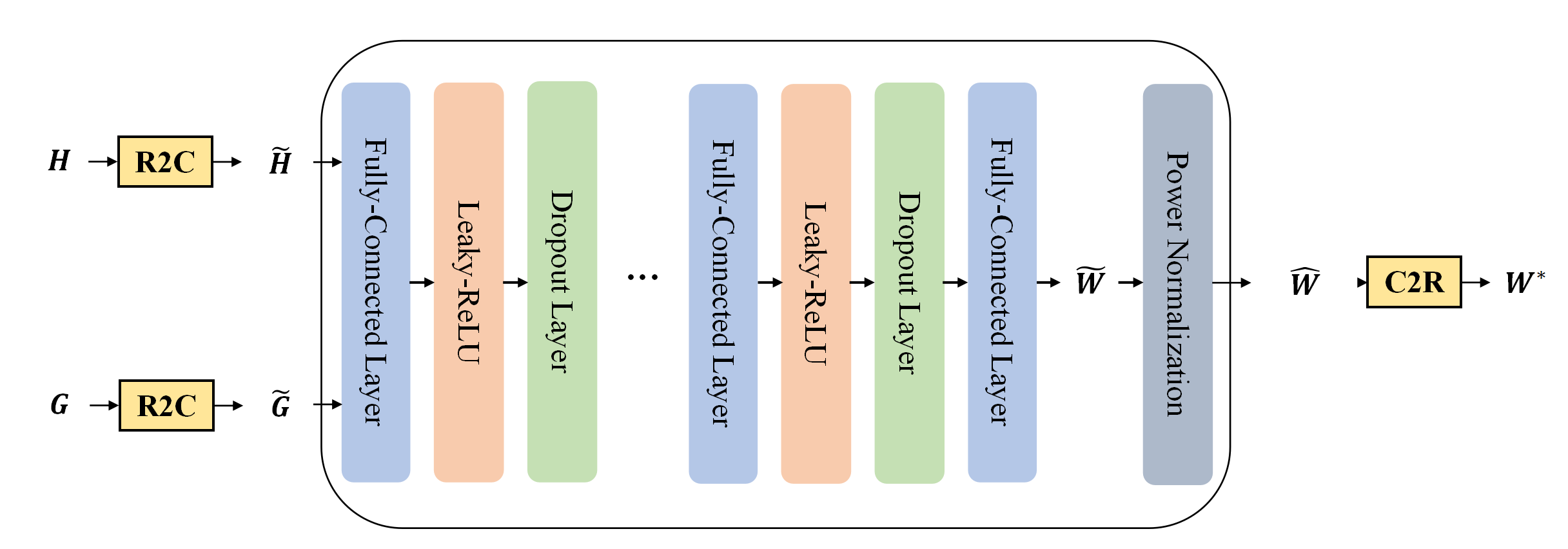}
\caption{The proposed beamforming DNN structure.}
\label{fig_DNN}
\end{figure}

Multilayer perceptrons (MLP) are recognized as universal funcion approximators \cite{hornik1989multilayer} and widely used in approaching complex nonlinear functions. As shown in Fig. \ref{fig_DNN},  we use a MLP network composed of the following units:
\subsubsection{Input layer}
For the DNN deployed at the $m$th BS, its first layer takes the communication channel $\mathbf{H}_m \in \mathcal{C}^{N_T \times MK}$ and sensing channel $\mathbf{G}_m \in \mathcal{C}^{N_T \times M}$ acquired at the local site as the input. As complex number operations are not supported by the current DNN software, the channel vector is transformed to real-valued coefficients before feeding it into the input layer through a $\mathbb{R}2\mathbb{C}$ block. Specifically, the communication channel matrix $\mathbf{H}_m$ is split into the real part $\Re(\mathbf{H}_m)$ and imaginary part $\Im(\mathbf{H}_m)$, which are then stacked to form the real-valued input vector $\tilde{\mathbf{H}}_m=[\Re(\mathbf{H}_m), \Im(\mathbf{H}_m)]$. Similarly, we process the sensing channel matrix $\mathbf{G}_m$ using the same pipeline and get $\tilde{\mathbf{G}}_m=[\Re(\mathbf{G}_m), \Im(\mathbf{G}_m)]$. The concatenated matrix $[\tilde{\mathbf{H}}_m, \tilde{\mathbf{G}}_m]$ is then flattened and fed into the first layer of the neural network for forward propagation. 

\IEEEpubidadjcol
\subsubsection{Hidden layers}
The MLP network contains $N_H$ fully-connected hidden layers with $d_H$ neurons in each layer. The depth of it impacts the nonlinearity and ability to extract feature information. Each hidden layer is followed by an activation layer and a dropout layer. As some elements of the beamforming matrix can be negative, the LeakyReLu \cite{maas2013rectifier} function is adopted as the activation function which provides a non-zero slope to negative values. The dropout layer with probability factor $\zeta=0.15$ can help to alleviate overfitting by randomly setting input units to zero. The process of hidden layers can be described as:
\begin{equation}
    \tilde{\mathbf{W}}=f([\tilde{\mathbf{H}}_m, \tilde{\mathbf{G}}_m];\mathbf{\omega})=\underbrace{f^{N_H-1}(f^{N_H-2}(...f^{1}([\tilde{\mathbf{H}}_m, \tilde{\mathbf{G}}_m])))}_{N_H\ \rm layers}.
\end{equation}
where $\mathbf{\omega}$ is the model parameters set.

\subsubsection{Output layer}
The output layer is of size $N_T\times K \times 2$ and followed by a normalization layer to scale the output so that the power constraint in \eqref{P:P1} can be satisfied, which can be denoted by
\begin{equation}
\label{eq:norm}
    \hat{\mathbf{W}} = \sqrt{\frac{P_T}{\text{tr}(\tilde{\mathbf{W}}\tilde{\mathbf{W}}^H)}}\tilde{{\mathbf{W}}}.
\end{equation}

Finally, we split the output $\hat{\mathbf{W}} \in \mathbb{R}^{N_T \times K \times 2}$ in the last dimension to obtain the real and imaginary parts, and process the output with a $\mathbb{C}2\mathbb{R}$ block, which recovers the complex beamformer $\mathbf{W}^*$ of size $N_T \times K$ by combining the real and imaginary parts in the following way
\begin{equation}
\label{eq:w}
    \mathbf{W}^* = \hat{\mathbf{W}}[:,:,1]+j\hat{\mathbf{W}}[:,:,2].
\end{equation}
where the two terms on the right-hand side are the first and second elements along the third dimension of $\hat{\mathbf{W}}$.

\begin{figure*}[htbp]
\centering
\includegraphics[scale=0.15]{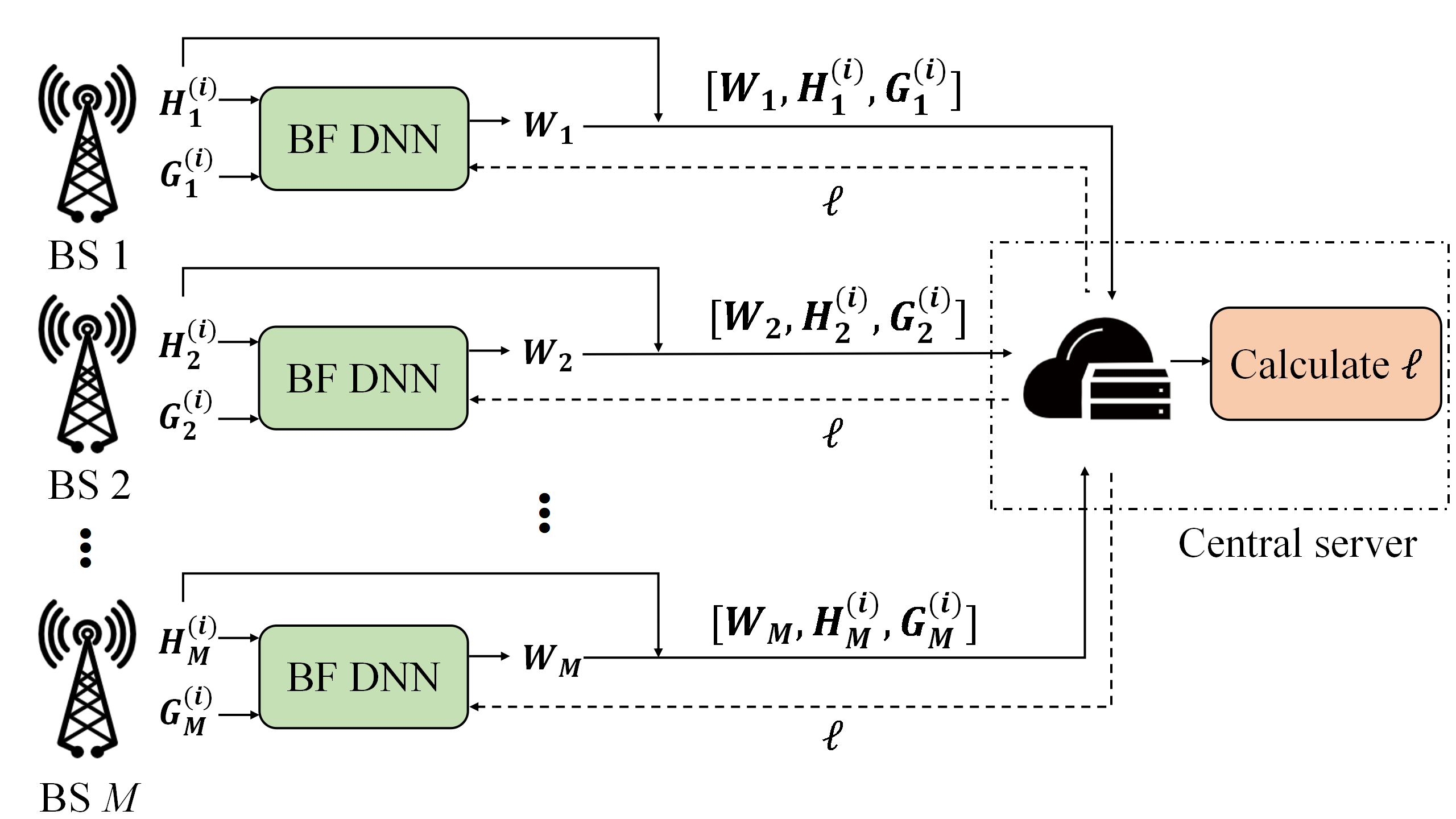}
\caption{VFL training framework.}
\label{fig_VFL}
\vspace{-5.0mm}
\end{figure*}

\IEEEpubidadjcol
\subsection{Learning-based Formulation and Training}
\label{VFL_training}
Due to the issue of local BSs lacking global information, it is difficult for them to directly calculate and optimize the local communication and sensing performance metrics for the intended users/targets as in \cite{liu2018toward, liuxiang2020joint, dong2020low, liu2021cramer, Hua10086626, li2024framework, sankar2023learning}. Traditional cooperative beamforming schemes \cite{meng2024cooperative1, babu2024precoding} require the collection of global channel information to design beamformers/precoders while managing inter-cell interference, which inevitably results in considerable communication overhead and increased latency during online beamforming stage. In contrast, our approach prioritizes minimizing information exchange and reducing the associated overhead. To achieve this, during the training phase, it is the central server that is responsible for calculating the global communication and sensing loss
\begin{equation}
\label{VFL_loss_com}
    \ell_c(\mathbf{W})=-\sum_{m=1}^{M}\sum_{k=1}^{K}\log_{2}(1+\gamma^c_{m,k}),
\end{equation}
\begin{equation}
\label{VFL_loss_sen}
    \ell_s(\mathbf{W})=-\sum_{m=1}^{M}\log_{2}(1+\gamma^s_{m}),
\end{equation}
where $\mathbf{W}=[\mathbf{W}_1,\mathbf{W}_2,..,\mathbf{W}_M] \in \mathbb{C}^{N_T \times KM}$ is the collection of locally designed beamforming matrices. Since the global loss functions \eqref{VFL_loss_com} \eqref{VFL_loss_sen} are defined as the negative sum communication rate and radar information rate, the global optimization problem becomes minimizing the weighted sum of (\ref{VFL_loss_com}) and (\ref{VFL_loss_sen})
\begin{equation}
    \begin{aligned}\label{P:P2}
        &\min_{\mathbf{[W_1,..,W_M]}} \quad \rho\ell_c(\mathbf{W}) + (1-\rho)\ell_s(\mathbf{W}) \\
        &\begin{array}{r@{\quad}r@{}l@{\quad}l}
        s.t. &\text{tr}(\mathbf{W}_m\mathbf{W}^H_m) \leq P_T,\quad \forall m.\\
        \end{array}
    \end{aligned}
\end{equation}
For the CUs, the complete feature space of the $i$th sample is the global channel information $\mathbf{H}^{(i)}=[\mathbf{H}^{(i)}_1, \mathbf{H}^{(i)}_2,...\mathbf{H}^{(i)}_M]$, with $\mathbf{H}^{(i)}_m$ being the $i$th partial feature available at BS $m$. Similarly, $\mathbf{G}^{(i)}=[\mathbf{G}^{(i)}_1, \mathbf{G}^{(i)}_2,...\mathbf{G}^{(i)}_M]$ denotes the $i$th complete feature space of sensing, which is the collection of local sensing channel samples. Under the VFL framework, each BS learns to design the local beamformer $\mathbf{W}^*_m, \forall m \in \mathcal{M}$ through the following steps.
\subsubsection{Forward propagation}
In the global round $T$, the local BSs use the $i$th local channel sample to design the beamforming matrices, the process can be given by 
\vspace{-2.0mm}
\begin{equation}
    \mathbf{W}_m=f([\mathbf{H}^{(i)}_m, \mathbf{G}^{(i)}_m];\mathbf{\omega}^{(T)}_m), 
    \vspace{-2.0mm}
\end{equation}
where $\mathbf{\omega}^{(T)}_m$ is the model parameter set of the $m$th BS at the $T$th global round.
 
\subsubsection{Local uploading}
Each BS uploads its independently designed beamforming matrices and the corresponding input channel sample $[\mathbf{H}^{(i)}_m,\mathbf{G}^{(i)}_m]$ to the central server.
\subsubsection{Features aggregation}
Upon receiving the designed beamformers and partial channel information from the BSs, the central server can aggregate them into the global channel information $\mathbf{H}^{(i)}$, $\mathbf{G}^{(i)}$ and calculate the communication loss (\ref{VFL_loss_com}) and sensing loss (\ref{VFL_loss_sen}). The global loss is then fed back to the local BSs for model updates.
\subsubsection{Backward propagation}
With the global loss information, each BS can perform individual backward propagation using the chain rule, propagating the error from the output layer back to the input layer of its own model. In this work, we adopt the stochastic gradient descent (SGD) algorithm \cite{amari1993backpropagation} to update the model parameters, which is given by
\vspace{-2.0mm}
\begin{equation}
\label{bp_VFL}
    \mathbf{\omega}^{(T+1)}_m=w^{(T)}_m-\eta\nabla\ell\big(\mathbf{W};\mathbf{\omega}^{(T)}_m\big),
    \vspace{-2.0mm}
\end{equation}
where $\eta$ represents the learning rate.

Fig. \ref{fig_VFL} shows the proposed VFL-based beamforming framework for cooperative beamforming design. Notably, the proposed network can be equivalently viewed as an overall multi-branch multi-output neural network, with each local model serving as one branch of the network and the server being the output layer. Through the backhual links, the central server guides the local models to control the ICI by implicitly providing the global channel information. The framework is summarized in \textbf{Algorithm} \textbf{\ref{alg:alg1}}.

\begin{algorithm}
\caption{Vertical Federated learning framework}\label{alg:alg1}
\begin{algorithmic}
\STATE \textbf{Offline Training:}
\STATE $\mathbf{Input}:\rm concatenated\ channel\ [\mathbf{H}^{(i)}_m, \mathbf{G}^{(i)}_m]$
\STATE $\mathbf{Output}: \rm beamformer\ \mathbf{W}^*$
\STATE $ \textbf{Initialize}\ \mathbf{\omega}^{(0)}_1, \mathbf{\omega}^{(0)}_2,..., \mathbf{\omega}^{(0)}_M $\;
\FOR{global round $T=1,2,...$}
\STATE \textbf{Forward Propagation:} each BS designs the respective beamforming matrix with local channel information;
\STATE \textbf{Features Aggregation:} the central server calculates the weighted global loss \eqref{VFL_loss_com} and \eqref{VFL_loss_sen} based on collected local information;
\STATE \textbf{Backward Propagation:} each BS updates local parameters with the global loss provided by the server \eqref{bp_VFL}.
\ENDFOR
\STATE \textbf{Online Beamforming:}
\STATE \hspace{0.5cm} $\mathbf{W}^*_m=f([\mathbf{H}^{(i)}_m, \mathbf{G}^{(i)}_m];\omega^*), \forall m \subset M$
\end{algorithmic}
\end{algorithm}

\IEEEpubidadjcol
\subsection{Complexity Analysis of VFL-based beamforming} 
\label{VFL_complexity}
Compared to traditional methods, in addition to the computational overhead during online beamforming stage, deep learning-based algorithms also involve the offline training costs. In the training phase, the total computational overhead can be generally represented as the sum of forward propagation and backward propagation, which is associated with the arithmetic operations in each fully connected layer and the data processing blocks. Specifically, the proposed MLP network consists of $N_H$ hidden layers, each with a dimension of $d_{H}$, while the input and output dimensions are related to the system configuration. Bsed on the Big-O notation, the computational complexity due to the linear transformation in the forward propagation stage can be denoted by $O((MK + M + K)N_Td_H+N_Hd^2_H)$. Since the activation layers involve element-wise operations, their complexity is given by $O((N_H+1)d_H)$. Besides, the input and output of the neural network are processed by the $\mathbb{R}2\mathbb{C}$ and $\mathbb{C}2\mathbb{R}$ blocks respectively, which results in the computational cost of $O((M+1)KN_T)$. Correspondingly, the computational cost of backward propagation is generally about twice that of forward propagation, as it involves similar matrix operations as in forward propagation, but with the additional gradient computation. Based on this, the total overhead associated with the computation during the training phase is given by $O(3((MK + M + K)N_Td_H+N_Hd^2_H))$.

\IEEEpubidadjcol
Furthermore, unlike traditional single-node deep learning, multi-node federated learning requires additional consideration of communication overhead during the training phase, in addition to the aforementioned computational costs. In a single training iteration of the VFL framework, the participating BSs need to upload their output matrices and the corresponding input channel samples to the central server for calculating the global loss after completing one forward propagation, which results in a total communication cost of $(M+1)KN_T$ for each local BS. 
As for the online deployment phase, the computational overhead is primarily attributed to forward propagation, which scales linearly with the number of cells $M$, communication users $K$ and transmit antennas $N_T$. As a comparison, the computational complexity of WMMSE is in the order of $O(L_\omega(K^2N^2_T+KN^3_T))$ according to \cite{hu2020iterative}, where $L_\omega$ denotes the number of iterations. This indicates the inefficiency of conventional optimization algorithms for practical use due to their complexity and reliance on multiple iterations for convergence, which limits parallelization. In contrast, neural networks leverage GPUs for parallel computing, greatly enhancing training and inference speeds.

\section{HFL-based Beamforming framework}
\label{HFL}
As discussed in the previous section, the proposed VFL framework eliminates the need for local channel information collection during online deployment, and addresses the issue of local BSs being unable to compute their local loss due to insufficient global channel information by calculating the global loss function at the central server. However, during the training phase, local channel samples still need to be uploaded to the central server at each iteration, and the training is conducted in a partially centralized manner. Besides, uploading large datasets incurs significant communication overhead and raises privacy concerns, which is is not conducive to the subsequent fine-tuning of the model. To further reduce the communication overhead in the absence of global channel information, we propose a fully decentralized beamforming framework based on HFL, which leverages the knowledge from all participating BSs while maintaining decentralized local datasets throughout both training and online deployment phases.

\begin{figure*}[htbp]
\centering
\includegraphics[scale=0.15]{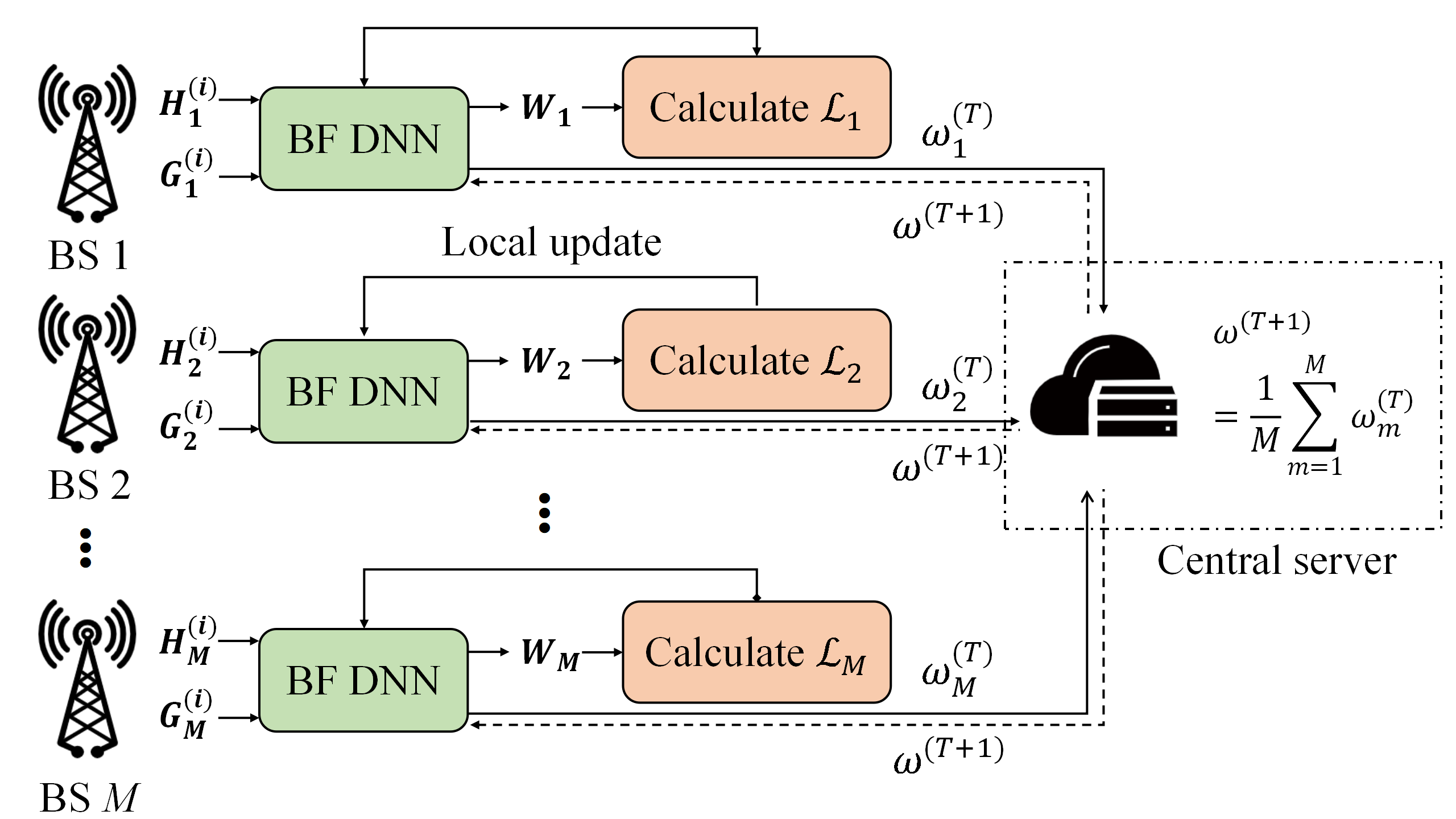}
\caption{HFL training framework.}
\label{fig_HFL}
\vspace{-5.0mm}
\end{figure*}

\IEEEpubidadjcol
\subsection{The proposed loss function}
As mentioned in the previous discussion, it is essential to understand the inherent complexity for calculating the ICI experienced by the local users and intended target at each BS due to the lack of global channel information. In conclusion, the problem (\ref{P:P1}) cannot be solved directly in a distributed manner.
Given the constraint of local channel information, we aim at designing a loss function for controlling the ICI that does not rely on direct access to global channel information from other BSs. Based on real-world conditions, we consider the following two cases.
\subsubsection{Interference-dominant scenario}
Interference-dominant scenario typically arises as the useful signal power increases, or when more users/targets move closer to the cell edges. At this point, the ICI becomes a more significant factor in affecting the network performance \cite{meng2024cooperative1}. Therefore, our goal is to minimize ICI to enhance both the user experience at the cell edges and the sensing performance. Following the workaround in \cite{babu2023multi}, we define the communication interference leakage (CIL)
\setcounter{equation}{18}
\begin{equation}
    \begin{aligned}\label{eq:leakc}
        \Phi^c_m(\mathbf{W}_m) &= \sum_{n\neq m}^{M}\sum_{i=1}^{K}|\mathbf{h}^H_{m,n,i}\mathbf{W}_m|^2.
    \end{aligned}
    \vspace{-2.0mm}
\end{equation}
Specifically, the aim is to minimize the interference caused to undesired receivers via generating the beamforming matrix from the null space spanned by the interfered channels. 
Similarly, each BS should also avoid the interference it causes to other BSs when sensing the target to enhance the network-level sensing performance, which can be accomplished by imposing constraints on the power of the sensing interference leakage (SIL)
\begin{equation}
\label{eq:leaks}
    \Phi^s_m(\mathbf{W}_m) = \sum_{n\neq m}^{M}|\mathbf{g}^H_{m,n}\mathbf{W}_m|^2.
    \vspace{-2.0mm}
\end{equation}
Employing the concept of CIL minimization allows us to ignore the ICI in the communication rate. Accordingly, the communication loss function $\mathcal{L}_c(\mathbf{W}_m)$ is formulated by combining the negative communication rate without considering communication ICI, and the CIL \eqref{eq:leakc} as the penalty term to be minimized. This is given in (\ref{eq:closs}), where $\alpha$ is the weighting factor affecting the importance of CIL management. Also, we formulate the sensing loss function $\mathcal{L}_s(\mathbf{W}_m)$ as the summation of negative radar information rate without sensing ICI, and the SIL \eqref{eq:leaks} term weighted by $\beta$, which is given in \eqref{eq:sloss}. It can be verified through simulations that optimizing loss function \eqref{eq:closs} and \eqref{eq:sloss}  can achieve a good performance when the interference caused to the $m$th cell is eliminated by other BSs, as shown in Section \ref{results}. Obviously, both \eqref{eq:closs} and \eqref{eq:sloss} can be calculated using local channel information, thus a fully decentralized training framework can be implemented.

\IEEEpubidadjcol
\subsubsection{Noise-dominant scenario}
In this case, the power of noise is much greater than that of the interference signal, which could be due to the low transmission power of the BS or because the users/targets are concentrated near the BSs, away from the cell edges. When the above conditions are satisfied, managing ICI becomes less important and BSs can focus on serving the intended users/targets by employing all available spatial resources to achieve the multiplexing gain.
At this stage, the weighting factors of interference leakage term in \eqref{eq:closs} and \eqref{eq:sloss} can be set to $\alpha=\beta=0$, and the corresponding local loss function at the $m$th BS is simplified into

\vspace{-5.5mm}
\setcounter{equation}{22}
\begin{small}
\begin{equation}
    \begin{aligned}\label{eq:closs2}
    \mathcal{L}_c(\mathbf{W}_m) &= -\sum_{k=1}^{K}\log_{2}\Big (1 + \frac{|\mathbf{h}^H_{m,m,k}\mathbf{w}_{m,k}|^2}{\sum_{l\neq k}^{K}|\mathbf{h}^H_{m,m,k}\mathbf{w}_{m,l}|^2 + \sigma^2_c}\Big),
    \end{aligned}
\end{equation}
\end{small}
\begin{small}
\begin{equation}
    \begin{aligned}\label{eq:sloss2}
    \mathcal{L}_s(\mathbf{W}_m) &= -\log_{2}\Big (1 + \frac{\sum_{k=1}^{K}|\mathbf{g}^H_{m,m}\mathbf{w}_{m,k}|^2}{\sigma^2_s}\Big),
    \end{aligned}
\end{equation}
\end{small}
which suggests the cooperative strategy degrades to the per-cell learning-based beamforming scheme as in \cite{huang2018unsupervised, sankar2023learning}.  

\newcounter{TempEqCnt} 
\setcounter{TempEqCnt}{\value{equation}} 
\setcounter{equation}{20} 

\begin{figure*}[!ht]
\normalsize
\begin{equation}
\label{eq:closs}
\begin{aligned}
        \mathcal{L}_c(\mathbf{W}_m) &= -\sum_{k=1}^{K}\Bigg(\log_{2}(1 +
        \frac{|\mathbf{h}^H_{m,m,k}\mathbf{w}_{m,k}|^2}{\sum_{l\neq k}^{K}|\mathbf{h}^H_{m,m,k}\mathbf{w}_{m,l}|^2 + \sigma_c^2}) - \alpha\sum_{n\neq 
        m}^{M}\sum_{i=1}^{K}|\mathbf{h}^H_{m,n,i}\mathbf{w}_{m,k}|^2 \Bigg). 
    \end{aligned}
\end{equation}

\begin{equation}
\label{eq:sloss}
\begin{aligned}
        \mathcal{L}_s(\mathbf{W}_m) &= -\Bigg(\log_{2}\Big (1 + \frac{\sum_{k=1}^{K}|\mathbf{g}^H_{m,m}\mathbf{w}_{m,k}|^2}{\sigma^2_s}\Big) - \beta \sum_{n\neq m}^{M}\sum_{k=1}^{K}|\mathbf{g}^H_{m,n}\mathbf{w}_{m,k}|^2\Bigg).
        \end{aligned}
\end{equation}
\hrulefill
\vspace{-5.0mm}
\end{figure*}

\IEEEpubidadjcol
\subsection{Learning-based Formulation and Training}
Based on the interference leakage-based loss functions \eqref{eq:closs} and \eqref{eq:sloss}, we aim at solving the following global optimization problem corresponding to fit the whole dataset $\mathcal{D}=\{ \mathcal{D}^{(m)}\}_{\forall m}$ across the BSs
\vspace{-2.0mm}
\setcounter{equation}{24}
\begin{equation}
    \begin{aligned} \label{P:P3}
         &\min_{\mathbf{W}} \quad \sum_{m=1}^{M}\Big (\rho\mathcal{L}_c(\mathbf{W_m}) + (1-\rho)\mathcal{L}_s(\mathbf{W_m})\Big )\\
         &\begin{array}{r@{\quad}r@{}l@{\quad}l}
         s.t. &\mathbf{tr}(\mathbf{W}_m\mathbf{W}^H_m) \leq P_T,\quad \forall m.
         \end{array}
    \end{aligned}
    \vspace{-2.0mm}
\end{equation}

With the HFL framework, the above problem can be solved in a decentralized manner with each BS solving the following local loss minimization problem
\vspace{-2.0mm}
\begin{equation}
    \begin{aligned} \label{P:P4}
        &\min_{\mathbf{W}_m} \quad \rho\mathcal{L}_c(\mathbf{W}_m) + (1-\rho)\mathcal{L}_s(\mathbf{W}_m)\\
        &\begin{array}{r@{\quad}r@{}l@{\quad}l}
        s.t. &\mathbf{tr}(\mathbf{W}_m\mathbf{W}^H_m) \leq P_T.
        \end{array}
    \end{aligned}
    \vspace{-2.0mm}
\end{equation}


Before the offline training phase, each BS collects the local channel samples using the method described in Section \ref{data_collect}. During the training stage, the participating BSs will receive the global model from the central server, and train the model for multiple iterations. In this work, we adopt the algorithm FedAvg \cite{mcmahan2017communication} to aggregate the global model, the details of offline training are as follows.
\subsubsection{Global broadcast}
At the beginning, the central server initializes a global model, and broadcasts it to all the participating BSs through the backhual link.
\subsubsection{Local training}
Upon receiving the newest global model $\mathbf{\omega}^{(T)}$ at the $T$th global iteration, each local BS $m$ performs forward propagation using the local channel dataset $\mathcal{D}^{(m)}$ to obtain a beamforming matrix $\mathbf{W}_m$ and computes the local loss function at the $t$th local iteration, which is the weighted sum of local communication loss (\ref{eq:closs}) and sensing loss (\ref{eq:sloss})
\begin{equation}
\label{eq:lossfunction}
    \begin{aligned}
        \mathcal{L}\big(\mathbf{W}_m;\mathbf{\omega}^{(T)(t)}_m\big)&=\rho\mathcal{L}_c\big(\mathbf{W}_m;\mathbf{\omega}^{(T)(t)}_m\big) \\
        &+ (1-\rho)\mathcal{L}_s\big(\mathbf{W}_m;\mathbf{\omega}^{(T)(t)}_m\big).
    \end{aligned}
\end{equation}
Then, the SGD algorithm is applied to update the local model 
\vspace{-2.0mm}
\begin{equation}
\label{bp}
    \mathbf{\omega}^{(T)(t+1)}_m=w^{(T)(t)}_m-\eta\nabla\mathcal{L}\big(\mathbf{W}_m;\mathbf{\omega}^{(T)(t)}\big).
    \vspace{-2.0mm}
\end{equation}
After multiple iterations of local training, the updated local model parameters $\mathbf{w}^{(T+1)}$ are obtained for model aggregation.
\subsubsection{Local uploading}
After the parallel local training, the updated model parameters $\mathbf{\omega}^{(T+1)}_m$, $m=1,2,...M$ are uploaded to the server through the backhaul link.
\subsubsection{Global aggregation}
Upon receiving the updated model parameters, the server performs global aggregation by averaging the received parameters \cite{mcmahan2017communication} to get the newest global model $\mathbf{\omega}^{(T+1)}$, then it decides whether to proceed with the next iteration of training depending on the current state.

The above procedure repeats until convergence is reached or a predefined number of iterations is completed. The key concept is that minimizing all local loss functions will lead to the minimization of the global loss function, thereby solving the problem (\ref{P:P3}). The HFL-based training framework is shown in Fig. \ref{fig_HFL}. Once the training is completed, the local BSs can use the trained network in real-time to design the beamforming matrix. Notably, during both training and deployment phases, the local datasets remain at the local site, which significantly reduces the communication overhead. The HFL framework is summarized in \textbf{Algorithm \ref{alg:alg2}}.

\begin{algorithm}
\caption{Horizontal Federated learning framework}\label{alg:alg2}
\begin{algorithmic}
\STATE \textbf{Offline Training:}
\STATE $\mathbf{Input}:\rm concatenated\ channel\ [\mathbf{H}^{(i)}_m, \mathbf{G}^{(i)}_m]$
\STATE $\mathbf{Output}: \rm beamformer\ \mathbf{W}^*$
\STATE $ \textbf{Initialize}\ \mathbf{\omega}^{(0)} $\;
\FOR{global round $T=1,2,...$}
\STATE the server broadcasts model parameters $\mathbf{\omega}^{(T)}$;
\FOR{local round $t=1,2,...$}
\STATE BSs minimize the local loss (\ref{eq:lossfunction}) and perform backward propagation (\ref{bp});
\ENDFOR
\STATE BSs upload model parameters to the server;
\STATE Global aggregation: $\mathbf{\omega}^{(T+1)}\leftarrow \frac{1}{M}\sum_{m=1}^{M}\mathbf{\omega}^{(T)}_m$
\ENDFOR
\STATE trained model $\omega^*$ feed back for local use;
\STATE \textbf{Online Beamforming:}
\STATE \hspace{0.5cm} $\mathbf{W}^*_m=f([\mathbf{H}^{(i)}_m, \mathbf{G}^{(i)}_m];\omega^*), \forall m \subset M$
\end{algorithmic}
\end{algorithm}

\begin{table*}[!t]
\caption{Comparison of communication overhead \label{tab:communication}}
\centering
\begin{tabular}{c|c|c}
\hline
Algorithm & Training Phase & Beamforming Phase\\
\hline
WMMSE[16] & \textbackslash & $M^2KN_T$ \\
\hline
VFL & $(M^2+M)TKN_T$ & \textbackslash \\
\hline
HFL & $2MBT$ & \textbackslash \\
\hline
\end{tabular}
\end{table*}

\begin{table*}[!t]
\caption{Comparison of computational complexity\label{tab:computation}}
\centering
\begin{tabular}{c|c|c}
\hline
Algorithm & Training Phase & Beamforming Phase\\
\hline
WMMSE[16] & \textbackslash & $O(L_\omega(K^2N^2_T+KN^3_T))$ \\
\hline
VFL & \multirow{2}{*}{$O(3(MK + M + K)N_Td_H+N_Hd^2_H)$} & \multirow{2}{*}{$O((MK + M + K)N_Td_H+N_Hd^2_H)$} \\
\cline{1-1}
HFL & & \\
\hline
\end{tabular}
\end{table*}

\IEEEpubidadjcol
\subsection{Complexity Analysis of HFL-based beamforming} 
\label{HFL_complexity}
Similar to the VFL-based method, the overhead of implementing HFL framework also consists of communication and computational costs during the training phase, as well as computational costs during the deployment phase. Since we are using the neural network of the same structure as in Section \ref{VFL}, the computational cost during the deployment phase for HFL-based framework, specifically the computation involved in the forward propagation, is identical to that of VFL-based framework and is given by $O((MK + M + K)N_Td_H+N_Hd^2_H)$. While in the offline phase, accounting for the backward propagation, the computational overhead arising from model training in HFL-based method can also be approximated as three times that of a single forward propagation. 

\IEEEpubidadjcol
In contrast to VFL-based method, the framework in this section significantly reduces communication overhead during the offline model training phase. Specifically, in a single training iteration of HFL, each BS only transmits the updated model parameters to the central server for global aggregation, instead of the training channel samples and designed beamforming matrices. As a result, the offline communication overhead arises from the model parameters broadcasting and uploading between local BSs and the server. This reduces the communication overhead as the memory occupied by the model parameters is much smaller compared to the training samples, and makes the framework to shift from feature aggregation to sample aggregation. Denoting the bits of the model parameters by $B$, the total bits to be uploaded and downloaded during distributed training is given by $2MBT$, where $T$ is the number of communication rounds to achieve a predefined performance $\epsilon$. According to \cite{li2019convergence}, the required number of communication rounds is given by $O\big[\frac{1}{\epsilon}\big((1+\frac{1}{K})EG^2+\frac{\Gamma+G^2}{E}+G^2\big)\big]$, where $E$ is the number of local training rounds and $G,\Gamma$ are problem-related constants. Besides, compared to Section \ref{VFL} where the global loss function is computed at the central server, in HFL framework, the loss function will be entirely based on locally available channel information and calculated in a fully decentralized manner by each local BS, which leverages the computational power of the participating BSs and offloads the computation workload from the server. In conclusion, this scheme achieves full decentralization during the model training phase, which significantly reduces communication overhead, facilitating future online model updates. For practical deployment, data compression techniques such as quantization and model pruning can be applied to further reduce the communication overhead. The comparison of communication overhead and computational complexity between the methods is summarized in Table. \ref{tab:communication} and \ref{tab:computation}.

\begin{figure*}[!t]
\centering
\subfigure[Communication rate $R_c$ versus SNRs.]{\includegraphics[width=2.5in]{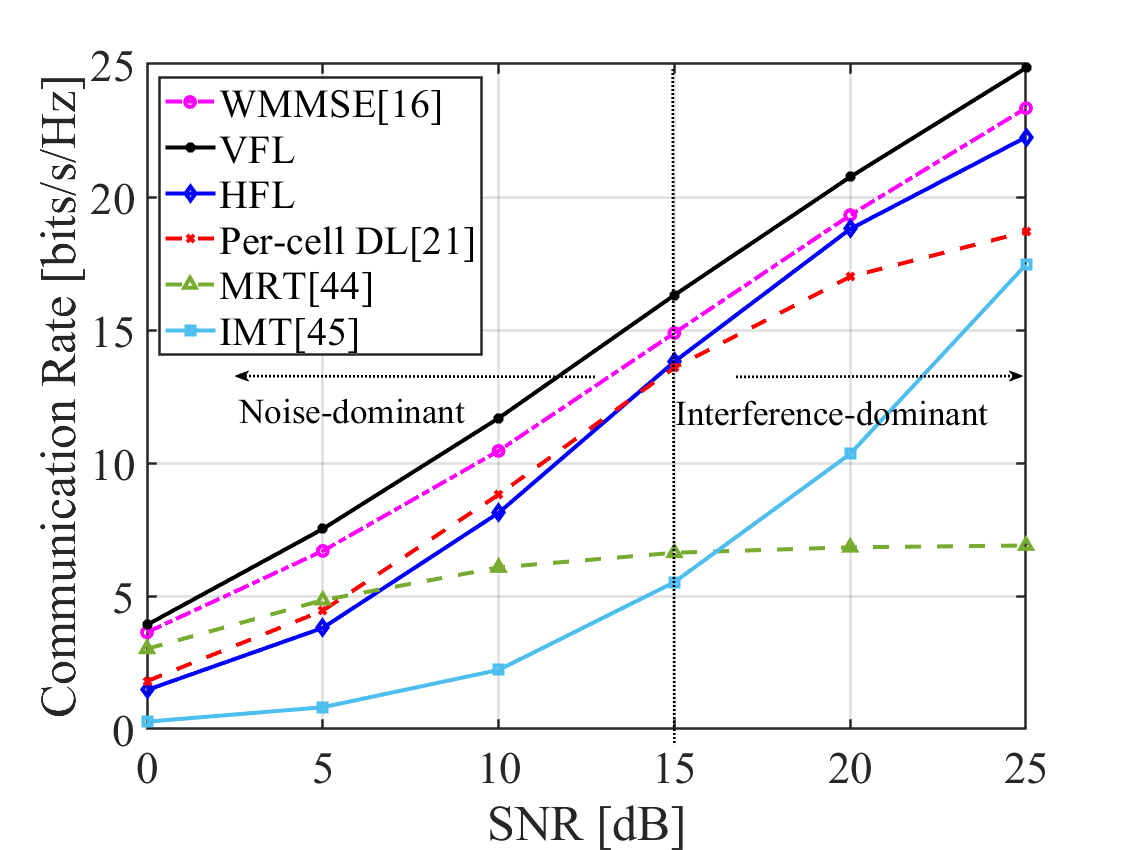}%
\label{fig_com}}
\hfil
\subfigure[Radar information rate $R_s$ versus SNRs.]{\includegraphics[width=2.5in]{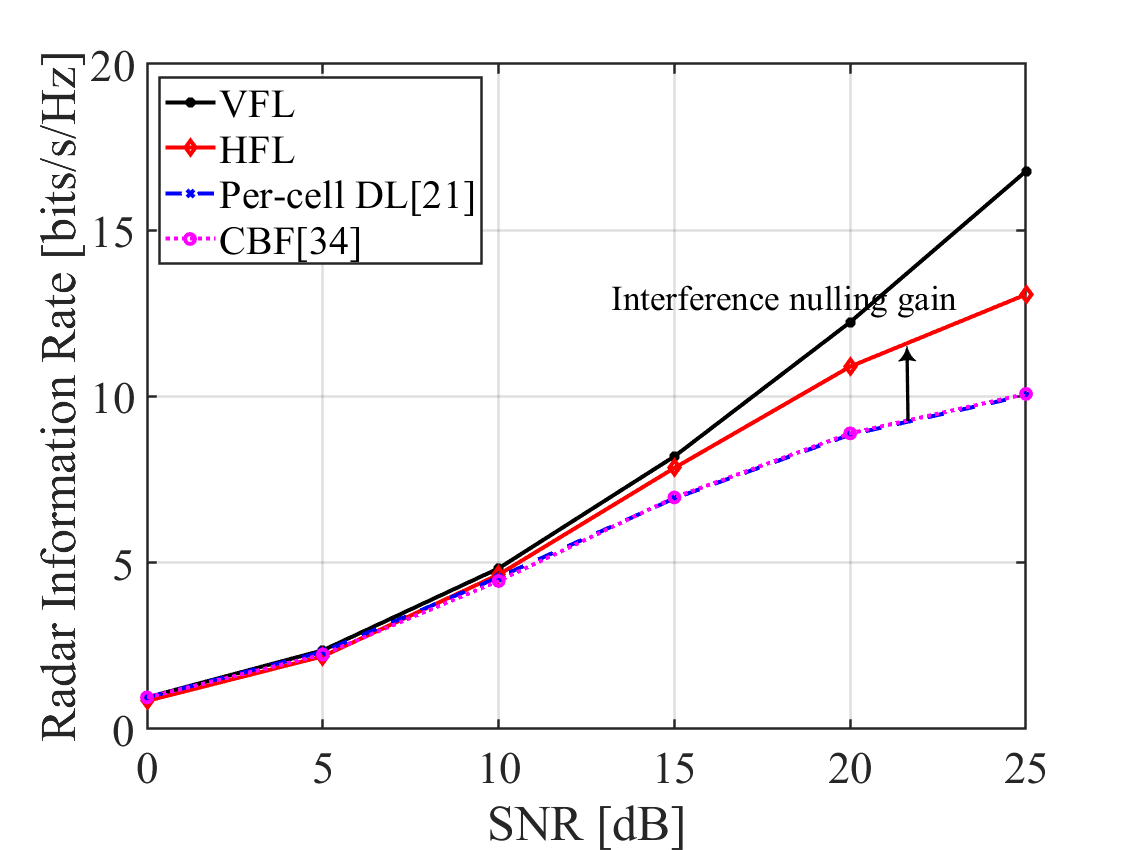}%
\label{fig_radar}}
\caption{Comparison of communication rate $R_c$ and radar information rate $R_s$ under different SNRs between the proposed methods and benchmarks.}
\end{figure*}

\IEEEpubidadjcol
\section{performance evaluation}
\label{results}
In this section, we illustrate the performance of the proposed FL-based beamforming frameworks via numerical simulations. The proposed frameworks are both implemented in Python 3.11.5 and Pytorch 2.1.0 on a PC with one NVIDIA RTX 3060 GPU and 8 Intel i7-11800 CPU cores.

\begin{table}[!t]
\caption{System parameters\label{tab:parameters}}
\centering
\begin{tabular}{|c|c|}
\hline
Parameter & Value\\
\hline
Cell radius & 500(m)\\
\hline
Communication path loss & 3.6\\
\hline
Sensing path loss & 2\\
\hline
Target RCS & 1.0($m^2$)\\
\hline
Num of hidden layers $N_H$ & 4\\
\hline
Num of neurons in hidden layers $d_H$ & 512\\
\hline
\end{tabular}
\end{table}

Table \ref{tab:parameters} summarizes the major hyper-parameters used for simulation. Unless otherwise mentioned, we adopted the following configurations adopted in the literature \cite{huang2018unsupervised, xia2019deep, kim2020deep}: the number of cells $M=3$, each BS is equipped with $N_T=6$ transmit antennas and $N_R=6$ receive antennas, serving $K=2$ users while sensing one point target. We produce channel samples by randomly generating the positions of users and targets. Specifically, the targets are assumed to be located within the range of $[-\pi/2, \pi/2]$ in the angular domain with respect to each BS, while the CUs are randomly distributed within the cell. Moreover, both communication channels and sensing ICI channels are assumed to be Rician fading channel with the rician factor being $3$.
Each local training dataset and test dataset contains 20,000 and 2000 channel samples respectively. During training, we apply the Adam optimizer \cite{kingma2014adam} to update the model parameters, and set weight decay factor to be $10^{-6}$ for further alleviating the risk of overfitting. 

\subsection{Evaluation of Communication Performance}
First, we evaluate the performance of the proposed methods by comparing the achievable communication rate with that of traditional methods under varying signal-to-noise ratios (SNRs). In this paper, we adopt the optimization-based scheme WMMSE \cite{shi2011iteratively}, the per-cell deep learning-based beamforming method \cite{huang2018unsupervised}, and the closed-form solutions maximum ratio transmission (MRT) \cite{fatema2017massive} and the interference minimizing transmission (IMT) \cite{zakhour2009distributed} as benchmarks. Fig. \ref{fig_com} shows the communication performance of our proposed methods when $\rho=0.99$. Throughout the increasing SNR levels, the VFL method consistently outperforms the benchmark solutions in all scenarios, while the HFL method approaches the performance of the WMMSE algorithm, with only $12\%$ performance gap comparing to VFL. This highlights the advantages of learning-based methods, as their computational complexity and latency are significantly lower than those of optimization-based approaches, while still surpassing the performance of closed-form solutions. It is worth noting that, the communication rates achieved by per-cell DL method and MRT scheme exhibit saturation at high SNRs (i.e., when the SNR exceeds approximately 15 dB in this case), as the network is becoming interference-limited in this regime. In contrast, the other schemes, which accounts for managing ICI, shows a significant performance improvement in interference-dominated scenarios, and this is referred to as the interference nulling gain. For example, when SNR $=25$dB, the HFL method achieves a $19\%$ improvement over the per-cell DL method by effectively constraining the interference leakage power in \eqref{eq:closs}. While with the aid of global channel information, the VFL method increases the performance improvement to $32\%$. As discussed in Section \ref{VFL_complexity} and \ref{HFL_complexity}, the tradeoff between the proposed VFL method and HFL method in terms of training overhead and online performance suggests that they can be tailored to suit different scenarios.

\begin{figure}[!t]
\centering
\includegraphics[width=2.5in]{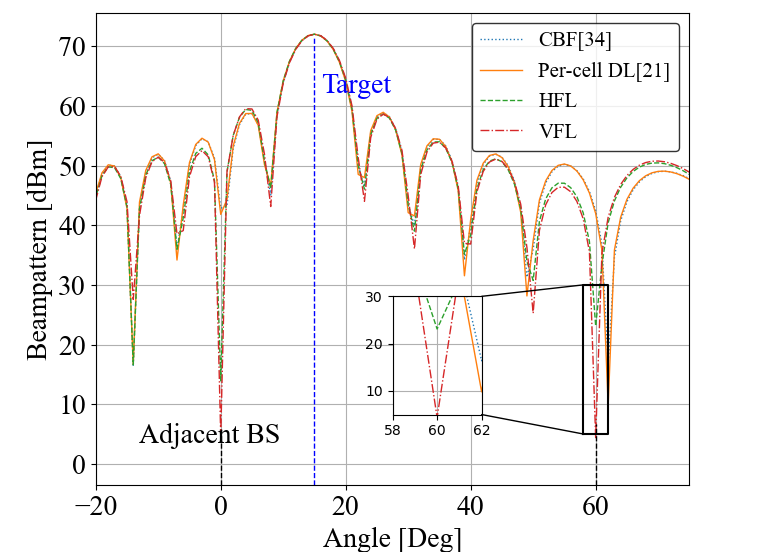}
\caption{Beampatterns for the scenario of point target at $15^{\circ}$.}
\label{fig_beampattern}
\end{figure}

\begin{figure*}[!t]
\centering
\subfigure[Communication rate $R_c$ versus number of transmit antennas.]{\includegraphics[width=2.5in]{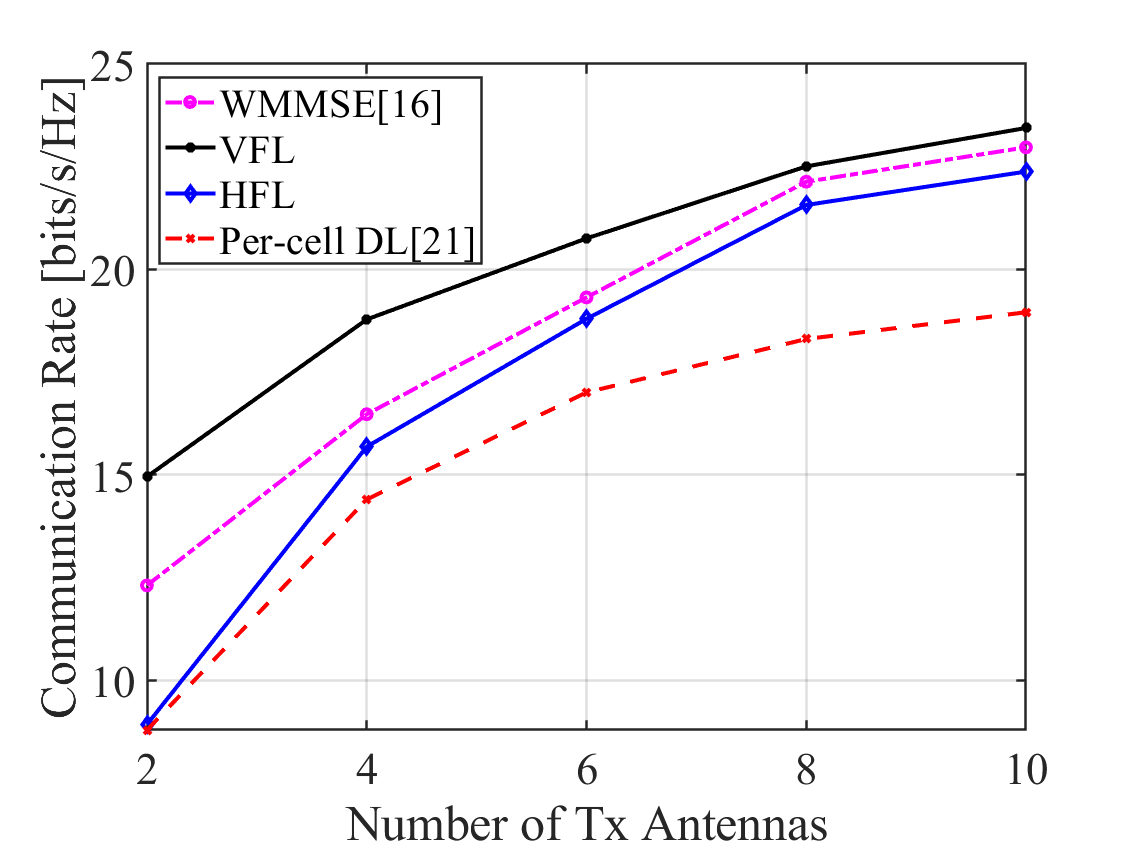}%
\label{fig_com_2}}
\hfil
\subfigure[Radar information rate $R_s$ versus number of transmit antennas.]{\includegraphics[width=2.5in]{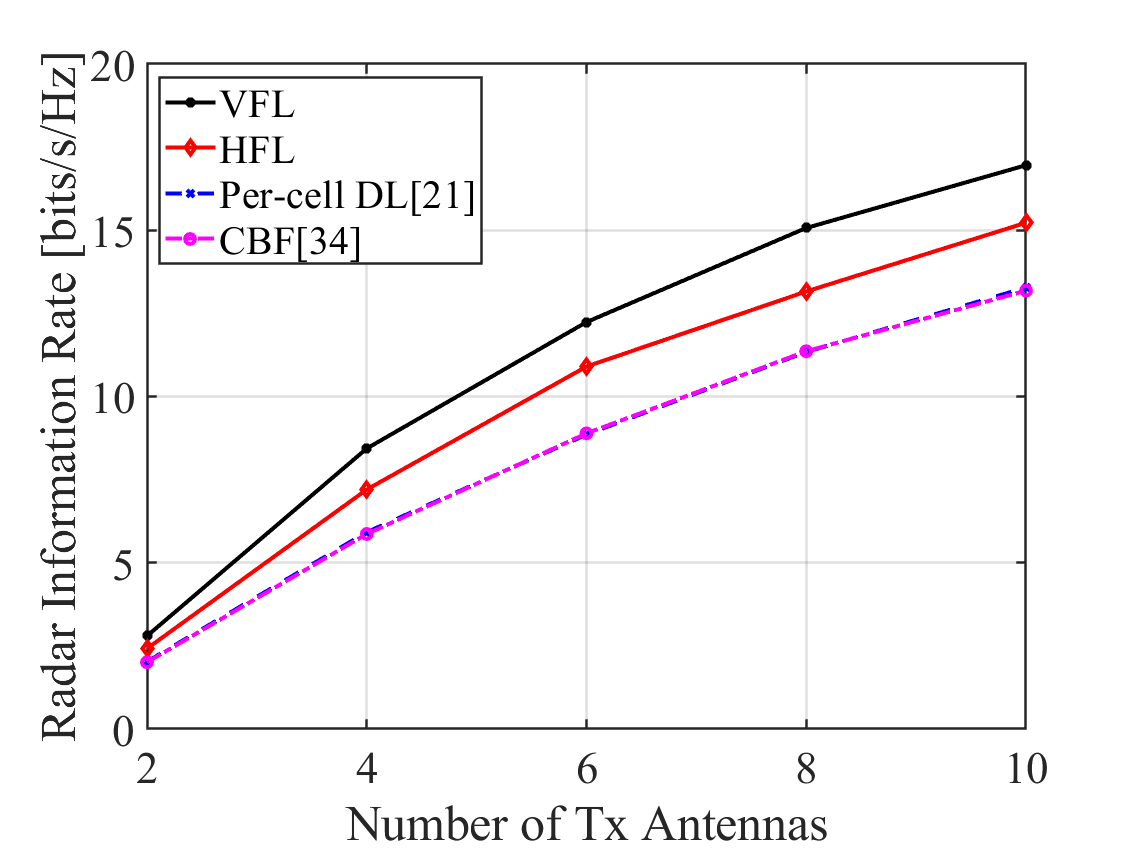}%
\label{fig_radar_2}}
\caption{Comparison of communication rate $R_c$ and radar information rate $R_s$ with respect to the number of antennas $N_T$ between the proposed methods and benchmarks.}
\end{figure*}

\subsection{Evaluation of Sensing Performance}
Fig. \ref{fig_radar} illustrates the sensing performance of the proposed methods when $\rho$ is $0.01$. We compare the achievable radar information rate of our methods with the closed-form scheme conjugate beamforming (CBF) \cite{demirhan2023cell}, where the transmit beamformer of the $m$th BS is towards the direction $\theta_m$ of the target to directly maximize the sensing SNR. In addition, we adopt the concept of unsupervised learning in \cite{huang2018unsupervised} and replaced its loss function from negative communication rate to negative radar information rate, using this as the benchmark for the per-cell DL-based method to highlight the impact of interference elimination on sensing performance. It can be observed that the proposed FL methods outperform CBF by up to $67\%$ and $30\%$ respectively in terms of the information rate as SNR increases, and the benefits brought by ICI elimination is more evident compared to that in communications in low SNRs. The result indicates that sensing ICI becomes a more dominant factor than noises and useful signal which suffers from round-trip pathloss affecting the sensing performance, which aligns with the conclusion drawn in \cite{meng2023network}. It is also worth noting that the curve of per-cell DL method overlaps with the curve of CBF. This is because when each BS is sensing only one target at the same time without considering the ICI, maximizing the illumination power will lead to the maximum radar information rate towards the target, at which point the principles of per-cell DL method and CBF are the same. Overall, these findings highlight the importance of incorporating ILM for optimizing both communication and sensing performance in multi-cell ISAC systems. 

The beampatterns obtained by different methods are shown in Figure. \ref{fig_beampattern}. We assume the serving BS is equipped with $N_T=N_R=16$ antennas and senses a point target located at $\theta=15^{\circ}$, while two adjacent BSs are located at $0^{\circ}$ and $60^{\circ}$. The result shows that the mainlobes of the transmit waveforms obtained by different methods align with the direction of the intended target, which can maximize the target illumination power. However, this cannot guarantee optimal network-level sensing performance, especially in the presence of ICI, as the unintended signal power leaking to surrounding BSs can disrupt their sensing. It is observed that the beampattern of per-cell DL method closely resembles that obtained by CBF method, which disregards the effect of sensing ICI towards adjacent BSs. Therefore, these two methods achieve the lowest sum radar information rate. In contrast, the beampattern of the VFL method exhibits rapid sidelobe attenuation at both $0^{\circ}$ and $60^{\circ}$, leading to less power leakage to the directions of adjacent BSs. By leveraging global channel information to train the local DNNs, this method maximally eliminates ICI to enhance the sensing performance. Similarly, by controlling interference leakage power using local channel information, the HFL method also achieves sensing interference nulling and the radar information rate comparable to that of VFL method.
\begin{figure*}[!t]
\centering
\subfigure[S\&C tradeoff under different SNRs.]{\includegraphics[width=2.5in]{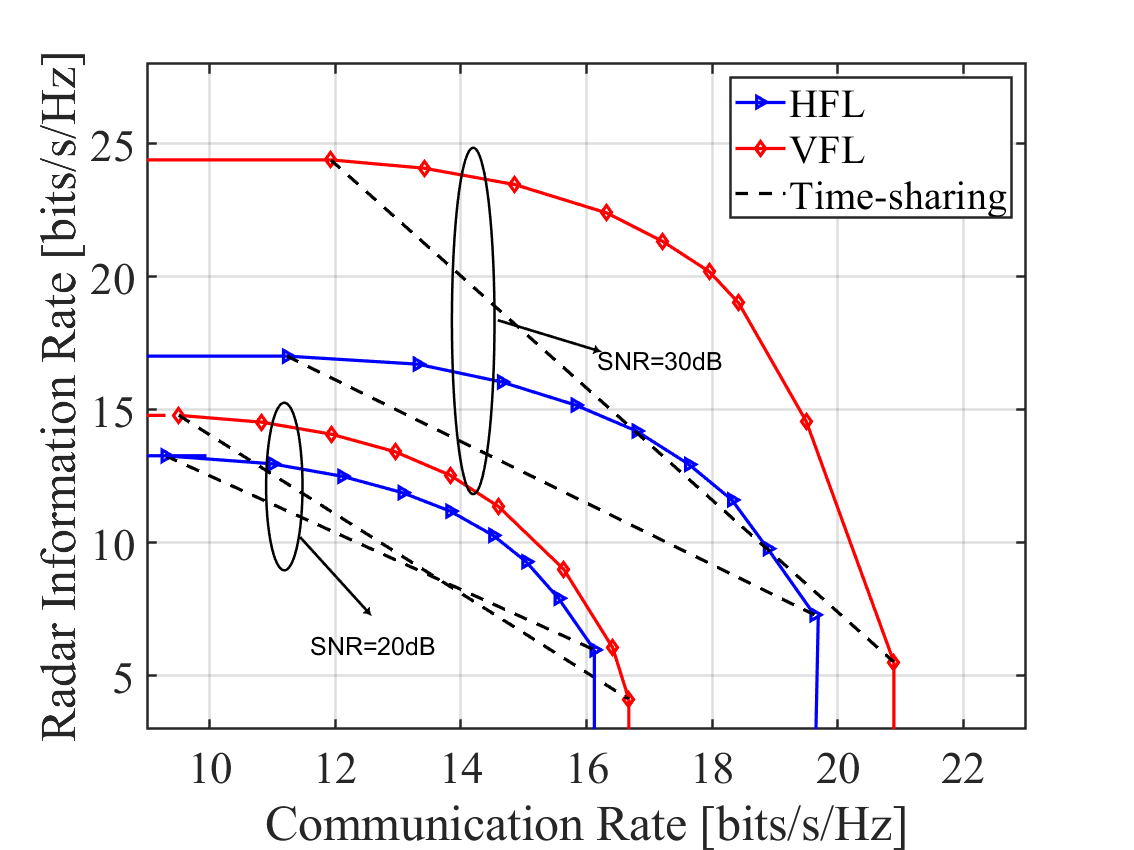}%
\label{fig_tf1}}
\hfil
\subfigure[S\&C tradeoff under different numbers of transmit antennas.]{\includegraphics[width=2.5in]{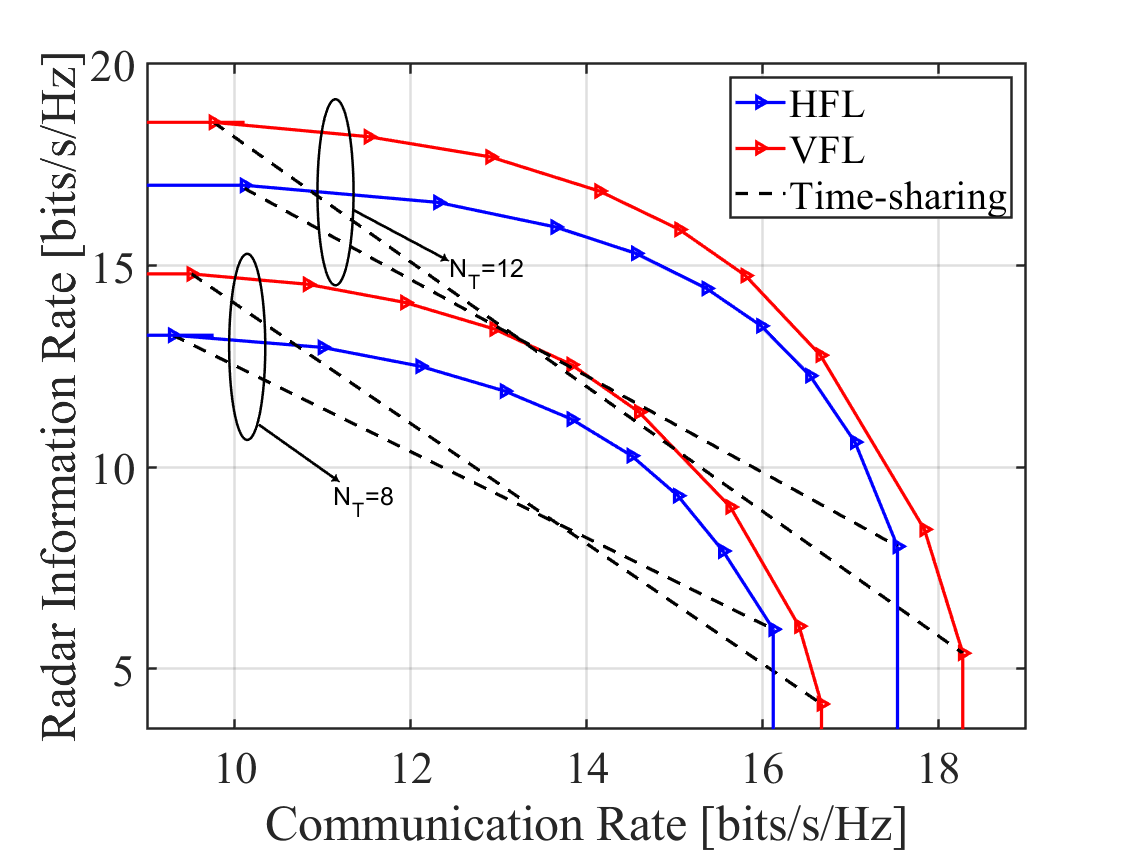}%
\label{fig_tf2}}
\caption{Tradeoff betweem communication rate $R_c$ and radar information rate $R_s$ when $\rho$ ranges from $0.1$ to $0.9$.}
\label{fig_tradeoff}
\end{figure*}

\subsection{Impact of the Number of Antennas}
In this section, we evaluate the S\&C performance of the proposed beamforming designs when varying the number of transmit antennas $N_T \in \{2,4,6,8,10\}$. Fig. \ref{fig_com_2} and Fig. \ref{fig_radar_2} show the achievable communication and radar information rate with respect to the number of antennas. This result demonstrates that our approaches exhibit good generalization capabilities across different antenna configurations. Besides, it is obvious that as the number of transmit antennas increases, the performance improvement of both VFL and HFL methods compared to per-cell DL method becomes more significant. This indicates that having access to global information and eliminating the ICI can greatly enhance both network-level S\&C performance. Therefore, we can conclude that when the spatial degrees of freedom are sufficient, eliminating ICI can lead to significant overall performance improvements for the network. By optimizing the use of available spatial resources, interference nulling becomes more impactful, especially in dense networks or scenarios with multiple users/targets at the cell edges.

\begin{figure*}[!t]
\centering
\subfigure[Effect of pruning to HFL method.]{\includegraphics[width=2.5in]{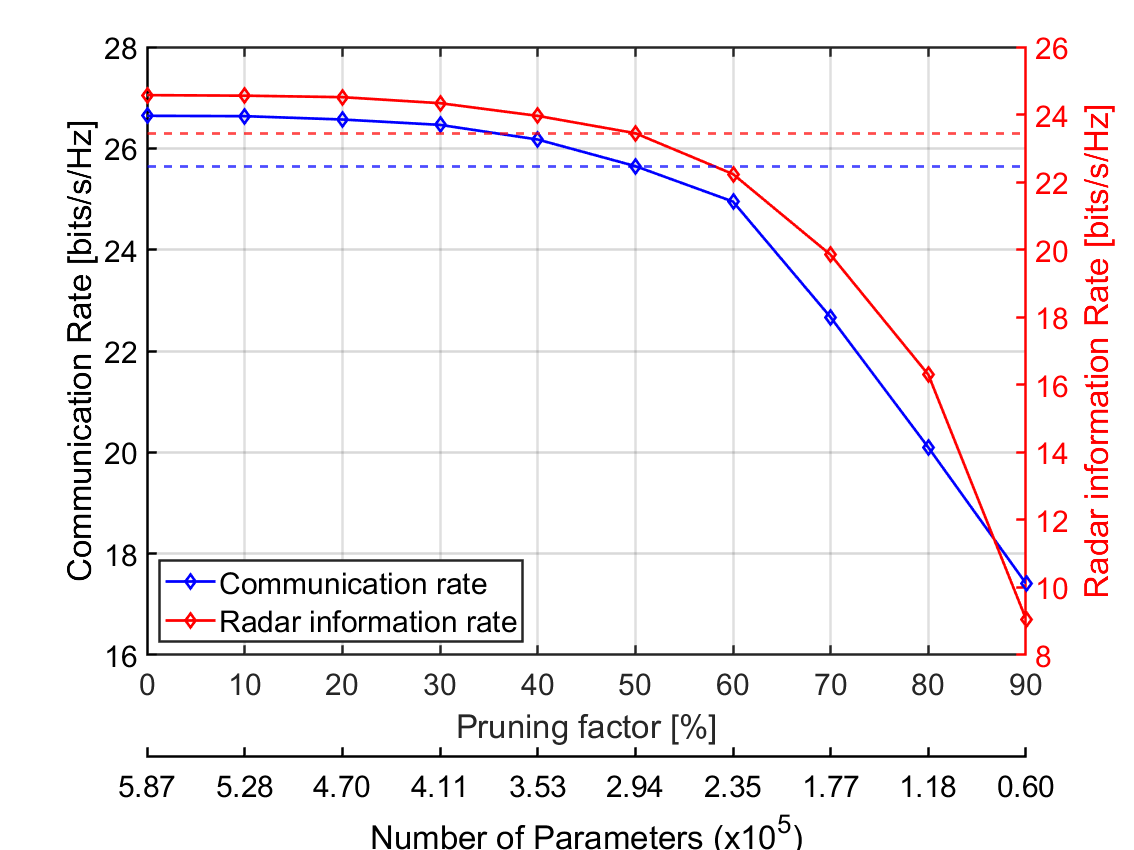}%
\label{fig_VFL_pruning}}
\hfil
\subfigure[Effect of pruning to VFL method.]{\includegraphics[width=2.5in]{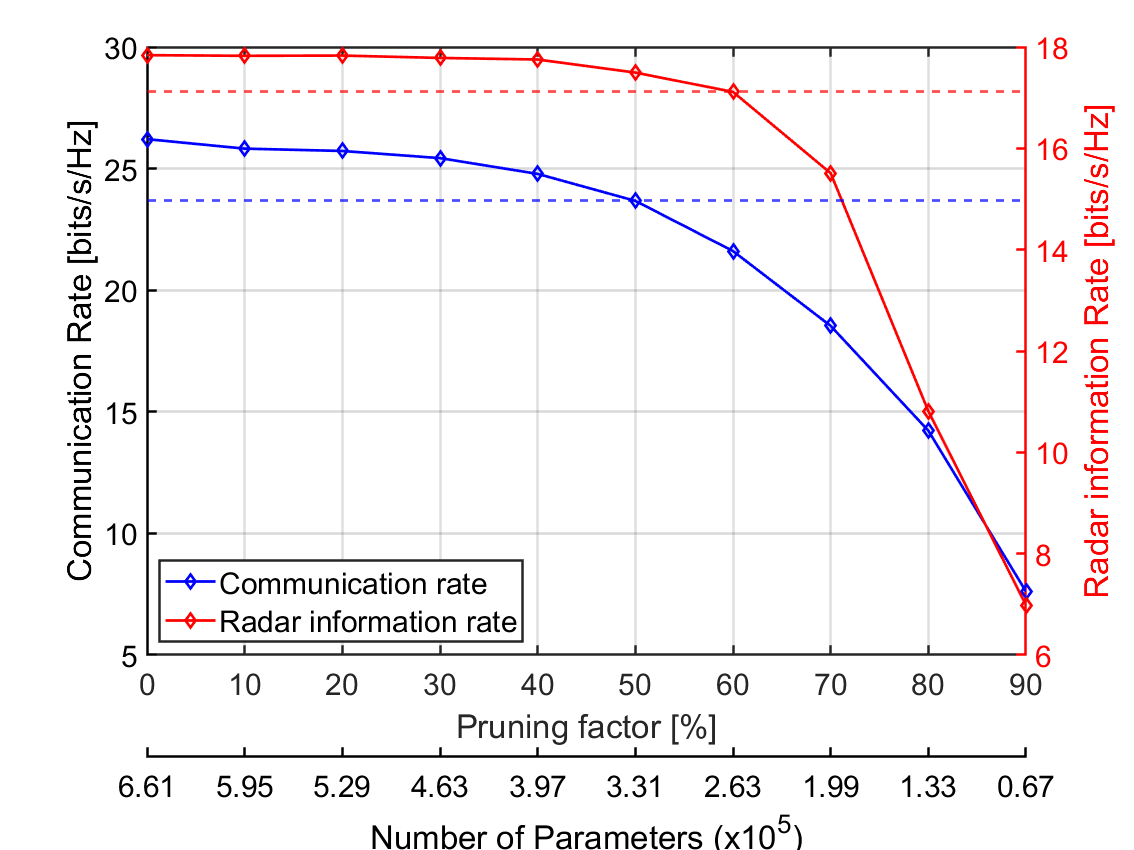}%
\label{fig_HFL_pruning}}
\caption{Performance vs complexity trade-off: effect of model pruning to the beamforming DNNs trained with different frameworks.}
\label{fig_pruning}
\end{figure*}

\subsection{Tradeoff Between Communication and Sensing}
The tradeoff profiles of the communication performance and sensing performance are shown in Fig. \ref{fig_tradeoff}. First, Fig. \ref{fig_tf1} illustrates the S\&C performance tradeoff under different SNRs, where each point represents the achievable communication rate $R_c$ and radar information rate $R_s$ as $\rho$ in (\ref{P:P1}) ranges from 0.1 to 0.9. Time-sharing scheme is introduced to evaluate the performance of our methods, which is achieved based on the two corner points. The boundary of the ($R_c$, $R_s$) region is viewed as the Pareto front constituted by the achieved performance tradeoff. From Fig. \ref{fig_tf1}, it can be observed that the boundaries of both HFL method and VFL method exhibit a gradual expansion as SNR increases, and the ($R_c$, $R_s$) region of proposed methods increase noticeably compared to time-sharing scheme at a higher SNR. For instance, with SNR$=20$dB, HFL method and VFL method outperform the time-sharing scheme by up to $29\%$ and $39\%$ respectively, in terms of the radar information rate under the same communication rate. When SNR rises to $30$dB, the improvements reach $33\%$ $50\%$ respectively. This suggests that the S\&C gain become more significant as more power is allocated and global channel information becomes available, leading to enhanced overall performance.

Second, Fig. \ref{fig_tf2} shows the effectiveness and generalization of proposed methods under different numbers of transmit antennas when SNR$=30$dB. Similar to the observed trend depicted in Fig. \ref{fig_tf1}, the ($R_c$, $R_s$) regions of both methods expand gradually compared to the times-sharing scheme when more transmit antennas are available. Specifically, for the HFL method, when fixing the communication rate, the achievable information rate is $31\%$ higher than that of the time-sharing scheme with $N_T=8$. While with $N_T=12$ deployed, this performance gap increases to $47\%$. Meanwhile, the VFL method can improve the radar information rate by up to $35\%$ and $51\%$ when $N_T=8$ and $N_T=12$ respectively as compared to the time-sharing scheme. This result demonstrates that our methods effectively leverage the available spatial DoFs, and the performance gain from eliminating ICI increases as the number of transmit antennas grows. The enhanced DoFs allow for more efficient interference management, leading to significant improvements in both network-level sensing and communication performance as antenna numbers scale up.

\subsection{A Scalable Perfromance vs Complexity Trade-off}
Considering the low latency and low overhead requirements in practical deployments, we apply model pruning \cite{han2015learning} technique to further improve the model’s efficiency. We compress the model size by removing less important weights and connections, with the amount removed controlled by the pruning factor. Fig. \ref{fig_pruning} shows that by selecting an appropriate pruning factor, the model's complexity and memory usage can be significantly reduced while maintaining performance. For instance, the horizontal lines in the figures indicate the point at which the model performance drops to approximately 95\% threshold, while both methods can maintain the performance above the threshold after removing 50\%-60\% of redundant nodes. At this point, the S\&C performance of both methods still surpass the closed-form solutions, while computational efficiency has been greatly improved. This allows for flexible adjustment of the pruning factor in practical deployments, facilitating a dynamic balance between performance and computational complexity according to specific operational requirements.

\section{Conclusion}
In this paper, we proposed two federated learning-based frameworks for the downlink beamforming design in multi-cell ISAC. We formulated the optimization problem to maximize the weighted sum of the system communication rate and radar information rate, and utilized the federated learning technique to find the optimal solutions efficiently. The first solution is based on the VFL framework, where the central server aggregates the channel information, and provides global information to train the local models. Considering the issue of deployment flexibility and training cost, we subsequently proposed the HFL method and the interference leakage-based loss function to enable a fully decentralized ICI management. The models trained under both frameworks can operate without global channel information. Through numerical simulations, we demonstrated that our solutions can achieve performance comparable to traditional centralized methods while offering significant improvements in computational efficiency and scalability, making it suitable for practical deployments. Our future work will involve advanced scenarios and individual target sensing requirements. 


\bibliographystyle{IEEEtran}
\bibliography{reference}

\end{document}